\documentclass[journal]{IEEEtran}
\usepackage{times,amsmath,amssymb,amsfonts,graphicx,epsfig,cite,psfrag}
\usepackage{algorithmicx,algorithm}
\usepackage{color}

\def\mubf{\boldsymbol \mu }

\def\Thetabf{\boldsymbol \Theta}

\def\abf{{\bf a}}

\def\Oc{{\cal O}}

\def\Tc{{\cal T}}

\def\defeq{~{\stackrel{\Delta}{=}}~}
\def\eg{{\it e.g.,\ \/}}

\def\ie{{\it i.e.,\ \/}}
\def\nn{\nonumber}

\def\w{\textrm{w}}

\newtheorem{theorem}{Theorem}
\newtheorem{lemma}{Lemma}
\newtheorem{proposition}{Proposition}

\begin{document}

\title{Residential Energy Storage Management with Bidirectional Energy Control }
\author{\IEEEauthorblockN {Tianyi Li, and Min Dong, \textit{Senior Member, IEEE}
\thanks{This work was supported by the National Sciences and Engineering Research Council of Canada under Discovery Grant RGPIN-2014-05181.}
\thanks{Tianyi Li was with Department of Electrical, Computer and Software Engineering, University of Ontario Institute of Technology, Oshawa, Ontario L1H 7K4, Canada. He is now with Huawei Technologies Co., Ltd, Markham, Ontario L3R 5A4, Canada (email: lty616@hotmail.com).}}
\thanks{Min Dong is with Department of Electrical, Computer and Software Engineering, University of Ontario Institute of Technology, Oshawa, Ontario L1H 7K4, Canada (email: min.dong@uoit.ca).}}
\markboth{IEEE Transactions on Smart Grid}{}
\maketitle

\begin{abstract}
We consider the residential energy storage management system with integrated renewable generation and the  availability of bidirectional energy flow from and to the grid through buying and selling. We  propose a real-time bidirectional energy control algorithm, aiming to minimize the net system cost from energy buying and selling as well as battery deterioration and storage inefficiency within a given time period, subject to the battery operational constraints and energy buying and selling constraints. We formulate the problem as a stochastic  control optimization problem. We then modify and transform this difficult problem into one that enables us to develop the real-time energy control algorithm through  Lyapunov optimization.
Our developed algorithm is applicable to arbitrary and unknown statistics of  renewable generation, load, and electricity prices. It provides a simple closed-form control solution based on only current system states, and requires a minimum complexity for real-time implementation. Furthermore, the   solution structure reveals how the battery energy level and energy prices  affect the energy flow direction and storage decision. The proposed algorithm possesses a bounded performance guarantee to that of the optimal non-causal $T$-slot look-ahead control policy.
Simulation shows the effectiveness of our proposed algorithm as compared with  alternative real-time and non-causal algorithms, and demonstrates the effect of selling-to-buying price ratio and battery inefficiency on the storage behavior and system cost.
\end{abstract}
\begin{IEEEkeywords}
Energy Storage, renewable generation, energy selling, home energy management, Lyapunov optimization, real-time control
\end{IEEEkeywords}

\section{Introduction} \label{sec:Introduction}

Energy storage and renewable energy integration are considered key solutions for  future power grid infrastructure and services to meet the fast rising energy demand and maintain energy sustainability \cite{dek10,CastilloGayme:Elsevier14}. For the grid operator, energy storage  can be exploited to shift energy across time to meet the demand and counter the fluctuation of intermittent renewable generation to improve grid reliability \cite{CastilloGayme:Elsevier14,SuGamal:TPS13,SunDongLiang:JSTSP14,SunLiangDongTaylor:TPS16,SunDongLiang:TSG16,ZhangGatsis_TSE13}. For the electricity  customer, local energy storage can provide means  for energy management to  control energy flow in response to the demand-side management signal and to reduce electricity cost. For example, dynamic pricing is one of main demand-side management techniques to relieve grid congestion \cite{Faruquietal:ElectrJ09,Wong&Sen&Ha&Chiang:JSAC2012}. Its effectiveness relies on the customer-side energy management solution to effectively control energy flow and demand in response to the price change. With local  renewable generation and energy storage introduced to residential and commercial customers, there are potentially greater flexibility in energy control  to respond to the dynamic pricing and demand fluctuation, as well as maximally harness the energy from renewable source to reduce electricity bills \cite{Wang&etal:TSG14,He&Zhang_TSG12,Li&Dong:JSAC15,Li&Dong:TSG16,Zhang&Schaar:JTSP14,MoghadamZhangMa:SGCOM14}.

With more local renewable generation at customers available, the utility retailer now allows customers to sell energy back to the utility at a price dynamically controlled by the utility in an attempt to harness energy from distributed renewable generation at customers and further improve stability and reliability \cite{ackermann2001overview}.   This means both renewable generation and previously stored energy, either bought from the grid or harnessed from the renewable source can be sold for profit by the customer. The ability to sell energy back to the grid enables bidirectional  energy flow between the energy storage system and the grid. This also gives the customer  a greater control capability to manage energy storage and usage    based on the dynamic pricing for both buying and selling. The repayment provides return for the  storage investment and further reduce the net cost at the customer. An intelligent energy  management solution exploring these features to    effectively manage storage and control the energy flow,  especially at a real-time manner, is  crucially needed to maximize the benefits.

Developing an effective energy management system faces many challenges. For the energy storage system itself, the renewable source is intermittent and   random, and its statistical characteristics over each day are often inheritably  time-varying, making it difficult to predict.
The benefit of storage, either for electricity supply or for  energy selling back,  also comes at the cost of battery operation that should not be neglected.
The bidirectional energy flow between the energy storage system and the grid under  dynamic pricing complicates  the energy control of the system when facing future uncertainty, and creates more challenges. More control decisions need to be made for energy flows  among storage battery, the grid, the renewable generation, and the load. The potential profit from energy selling under the unpredictable pricing complicates the control decisions  in terms of when and how much to sell, store, or use. Moreover, the battery capacity limitation further makes the control decisions coupled over time and difficult to
optimize. In this paper, we aim to develop a real-time energy control solution that  addresses these challenges and effectively reduces the net system cost at minimum required knowledge of unpredictable system dynamics.


\subsection{Related Works}
Energy storage has been considered at the power grid operator or aggregator to combat the fluctuation of renewable generation, with many works in literature on storage control and assessment of its role in renewable generation \cite{CastilloGayme:Elsevier14}, for power balancing with fixed load \cite{SuGamal:TPS13,SunDongLiang:JSTSP14} or flexible load control \cite{ZhangGatsis_TSE13,SunDongLiang:TSG16}, and for phase balancing \cite{SunLiangDongTaylor:TPS16}. Residential energy storage systems to reduce electricity cost have been considered without renewable \cite{Urgaonkar&Neely:SIGMETRICS2011} and with renewable integration \cite{He&Zhang_TSG12,RahbarXu:TSG15,Wang&etal:TSG14,Huang&Walrand&Ramchandran:SmatGridComm12,
Sergio&Ming&Pan&Yong_TSG13,Li&Dong:ICASSP2013,Li&Dong:acssc2013,Li&Dong:SmartGridComm14,Li&Dong:JSAC15,Li&Dong:TSG16,Qinetal:TSG16}. Only energy buying was considered in these works.  Among them, off-line storage control  strategies for dynamics systems are proposed \cite{He&Zhang_TSG12,RahbarXu:TSG15,Wang&etal:TSG14}, where combined strategies of load prediction and day-ahead scheduling on respective large and small
timescales are proposed. The knowledge of load statistics  and renewable  generation are known ahead of time, while no battery operational cost are considered.

Real-time energy storage management amid unknown system dynamics is much more challenging. Assuming known distributions of system dynamics (\eg load, renewable generation, and prices), the storage control problem is formulated as a Markov Decision Process (MDP) and solved numerically using Dynamic Programming \cite{Zhang&Schaar:JTSP14,RahbarXu:TSG15}. However, this method suffers from high computational complexity to be implementable for practical systems.  In addition, due to unpredictable nature of system dynamics, the required statistics are difficult to acquire or predict in practice. Without the statistical knowledge of system dynamics, Lyapunov optimization technique \cite{book:Neely} has been employed to develop real-time control strategies in
\cite{Li&Dong:JSAC15,Li&Dong:TSG16,Huang&Walrand&Ramchandran:SmatGridComm12,Sergio&Ming&Pan&Yong_TSG13,Li&Dong:ICASSP2013,Li&Dong:acssc2013}. For independent and\ identically distributed  or stationary system dynamics (pricing, renewable, and load), energy control algorithms are proposed in \cite{Huang&Walrand&Ramchandran:SmatGridComm12,Sergio&Ming&Pan&Yong_TSG13} without considering battery operational cost, and in  \cite{Li&Dong:ICASSP2013}  with battery charging and discharging operational cost considered. All the above works  aim to minimize the long-term average system cost. A real-time energy control algorithm to minimize the system cost within a finite time period is designed in \cite{Li&Dong:JSAC15} for arbitrary system dynamics. Furthermore, joint storage control and flexible load scheduling is considered in \cite{Li&Dong:TSG16} where the closed-form sequential solution was developed  to minimize the system cost while meeting the load deadlines.

The idea of  energy selling back or trading  is considered in \cite{Kimetal:JSAC13,Mediwaththe&etal:TSG17,Huang&Mao&Nelms:J_SmartGrid2014}, where    \cite{Kimetal:JSAC13,Mediwaththe&etal:TSG17} focus on demand-side management via pricing schemes using game approaches for load scheduling among customers, and \cite{Huang&Mao&Nelms:J_SmartGrid2014} considers a microgrid operation and supply. In addition, although not explicitly modeled, the system considered in \cite{Qinetal:TSG16} can be generalized to include energy selling under a simplified model, provided that  buying and selling prices are constrained such that the overall cost function is still convex. All these works consider the grid level operation  and the cost associated with it, and use a simple battery storage model without  considering degradation or operational cost. Since the consumers may prefer a cost saving management solution in a customer-defined time period, and system dynamics may not be stationary, it is important to provide a cost-minimizing  solution to meet such need.  To the best of our knowledge, there is no such existing   bidirectional energy  management solution with energy selling-back capability. In addition, most existing works ignore battery inefficiency in charging and discharging, which results in energy loss that affects the storage behaviors and should be taken into account in the energy storage control design.

\subsection{Contributions}
In this paper, we consider a residential energy storage management system with integrated renewable generation and the availability of bidirectional energy flow from and to the grid through  buying and selling. We develop a real-time bidirectional energy  control algorithm,  aiming to minimize the net system cost in a finite time period, subject to the battery operational constraints and energy buying and selling constraints. In considering the system cost, we include the energy storage cost by carefully modeling both battery  operational cost and the inefficiency associated with charging/discharging activities. The system dynamics, including renewable generation, buying/selling electricity price, and the customer load, can have arbitrary distributions which may not be stationary, and are unknown to us.

We formulate the net system cost minimization as a stochastic optimization problem over a finite time horizon. The interaction of storage, renewable, and the grid as well as the cost associated with energy buying/selling and battery operation for storage complicate the energy control decision making and optimization over time. To tackle this difficult  problem, we use special techniques to modify and transform the original problem into one that we are able to apply Lyapunov optimization to develop a real-time  algorithm for the control solution. Our developed real-time    algorithm provides a simple closed-form energy control solution, which only relies on  the current battery energy level, pricing, load, and renewable generation. It has a minimum complexity for real-time implementation. In addition, the closed-form expression  reveals how the battery energy level and prices  affect the decisions of energy buying and selling, storage and usage of the battery, and the priority order  for storing or selling energy  from multiple energy sources. Through analysis, we show that the performance of our proposed real-time algorithm is within a bounded gap to that of the optimal $T$-slot look-ahead solution which has full  system information available ahead of time. The algorithm is also shown to be asymptotically optimal as the battery capacity and the time duration go to infinity.   Simulation results demonstrate the effectiveness
of our proposed algorithm as compared with alternative real-time or non-causal control solutions. Furthermore, we provide simulation studies  to  understand the effects of bidirectional energy control, the selling and buying price ratio, and battery efficiency on the energy storage behavior and the system cost.

\subsection{Organization and Notations} The rest of this paper is organized as follows. In Section~\ref{sec:model}, we describe the bidirectional energy storage and management system model. In Section~\ref{sec:FHA}, we formulate the  stochastic energy control optimization problem  and develop an approach to transform it  for the real-time control design. In Section~\ref{sec:RT alg}, we present our real-time energy control algorithm. In Section~\ref{sec:PA}, we analyze the performance of our algorithm. Simulation results are provided in Section~\ref{sec:sim}, and followed by conclusion in Section~\ref{sec:conclusion}.

\emph{Notations}:
The main symbols used in this paper are summarized in Table \ref{tab:num}.
\begin{table}[t]
\renewcommand{\arraystretch}{1.4}
\centering
\caption{List of main symbols}\label{tab:num}
\begin{tabular}{p{0.7cm}| p{7.1cm}}
\hline\hline
$W_t$ & customer load at time slot $t$\\
\hline
$S_t$ & renewable generation at time slot $t$ \\
\hline
$S_{w,t}$ &  portion of renewable energy serving load $W_t$ at time slot $t$\\
\hline
$S_{c,t}$ & portion of renewable energy sold to grid at time slot $t$\\
\hline
$S_{s,t}$ & portion of renewable energy stored to battery at time slot $t$\\
\hline
$E_t$ & energy bought from conventional grid at time slot $t$ \\
\hline
$Q_t$ & portion of $E_t$ stored into battery at time slot $t$\\
\hline
$F_{d,t}$ & amount of energy from battery serving load $W_t$ at time slot $t$\\
\hline
$F_{s,t}$ & amount of energy from battery sold to grid at time slot $t$\\
\hline
$P_{b,t}$ & energy unit buying price from conventional grid at time slot $t$\\
\hline
$P_{s,t}$ & energy unit selling price to conventional grid at time slot $t$\\
\hline
$B_t$ & battery energy level at time slot $t$\\
\hline
$x_{e,t}$ & entry cost for battery usage at time slot $t$:\newline  $x_{e,t}= 1_{R,t}C_{\textrm{rc}}+1_{D,t}C_{\textrm{dc}}$\\
\hline
$x_{u,t}$ &  net change of battery energy level  in time slot $t$:\newline $x_{u,t}=
\left|Q_t+S_{r,t}-D_t\right|$\\
\hline
$\abf_t$ & control actions at time slot $t$:\newline $\abf_t\triangleq[E_t,Q_t,F_{d,t},F_{s,t},S_{c,t},S_{s,t}]$\\
\hline
$\mubf_t$ & system inputs  at time slot $t$: $\mubf_t\triangleq[W_t,S_t,P_{b,t},P_{s,t}]$\\
\hline
$\overline{x_e}$ & average entry cost for battery usage over  $T_o$ slots\\
\hline
$\overline{x_u}$ & average net change of battery energy level over  $T_o$  slots\\
\hline
$\overline{J}$ & average cost of buying energy from the grid over $T_o$ slots \\
\hline
$C(\cdot)$ & average usage cost function  of the battery\\
\hline
$T_o$ & time period in slots considered for system cost minimization \\
\hline
$\eta_c$ & battery charging efficiency factor\\
\hline
$\eta_d$ & battery discharging efficiency factor\\
\hline
$R_{\max}$ & maximum charging amount into the battery\\
\hline
$D_{\max}$ & maximum discharging amount from the battery\\
\hline
$\Gamma$ & $\max\{\eta_cR_{\max},D_{\max}/\eta_d\}$ \\
\hline
$B_{\min}$ & minimum energy required in battery\\
\hline
$B_{\max}$ & maximum energy allowed in battery\\
\hline
$C_{\textrm{rc}}$ & entry cost for battery usage due to each charging activity\\
\hline
$C_{\textrm{dc}}$ & entry cost for battery usage due to each discharging activity\\
\hline
$\Delta_a$ & desired change amount of battery energy level in $T_o$ slots\\
\hline\hline
\end{tabular}
\end{table}

\section{System Model} \label{sec:model}
\begin{figure}[t]
\centerline{
\begin{psfrags}
    \psfrag{C1}{\normalsize $(P_{b,t})$}
    \psfrag{C2}[c]{\normalsize $(P_{s,t})$}
    \psfrag{P}{\normalsize ${E_t}$}
    \psfrag{S}{\normalsize ${S_t}$}
    \psfrag{PQ}{\normalsize ${E_t-Q_t}$}
    \psfrag{FS}{\normalsize ${F_{s,t}+S_{s,t}}$}
    \psfrag{Q}{\normalsize ${Q_t}$}
    \psfrag{S1}{\normalsize ${S_{w,t}}$}
    \psfrag{S2}{\normalsize ${S_{c,t}}$}
    \psfrag{S3}{\normalsize ${S_{s,t}}$}
    \psfrag{F1}{\normalsize ${F_{d,t}}$}
    \psfrag{F2}{\normalsize ${F_{s,t}}$}
    \psfrag{W}{\normalsize ${W_t}$}
    \psfrag{X}{\normalsize $+$}
    \psfrag{K}{\normalsize $+$}
    \psfrag{O}{\normalsize $-$}
    \psfrag{Y}{\normalsize $-$}
    \psfrag{Renew.}{\normalsize RG}
    \psfrag{Grid}{\normalsize CG}
    \psfrag{Battery}{\small Battery}
    \psfrag{Contrl}{\small Control}
    \psfrag{User}{\normalsize Load}
    \includegraphics[width=0.9\linewidth]{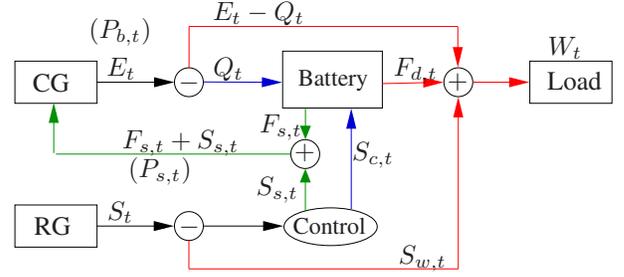}
\end{psfrags}
}
\caption{An ESM system with RG and bidirectional energy flow from/to CG.}
\label{fig:sellingback_model}
\end{figure}

We consider a residential-side energy storage  management (ESM) system as shown in Fig.~\ref{fig:sellingback_model}. The system contains an energy storage battery which is connected to an on-site renewable generator (RG) and the conventional grid (CG). Energy can be charged into the battery from both the RG and the CG, discharged from the battery for customer electricity demand, or sell back to the CG. Both the RG and the CG can   directly supply energy to the customer. We assume the ESM system operates in discrete time slots with $t\in \{1,2,\cdots\}$, and all energy control operations are performed per time slot $t$.
\subsubsection{RG} Let $S_t$ be the amount of energy harvested from the RG at time slot $t$. Due to the uncertainty of the renewable source, $S_t$ is random. We assume no prior knowledge of $S_t$ or its statistics.
 Let  $W_t$ be the customer load at time slot $t$. We assume a priority of using $S_t$  to directly serve $W_t$. Let $S_{w,t}$ be this portion of $S_t$ at time slot $t$. We have $S_{w,t}=\min\{W_t,S_t\}$. A controller will determine whether the remaining portion, if any,  should be stored into the battery and/or sold back to the CG. We denote the stored amount and sold amount by $S_{c,t}$ and $S_{s,t}$, respectively, satisfying
\begin{align}
S_{c,t}+S_{s,t}\in[0,S_t-S_{w,t}]\label{eqn:Solar S2 S3 bounds}.
\end{align}
\subsubsection{CG} The customer can buy energy from or sell energy to the CG at real-time unit  buying price $P_{b,t}\in[P_{b}^{\min}, P_{b}^{\max}]$ and selling price $P_{s,t} \in [P_{s}^{\min}, P_{s}^{\max}]$, respectively. Both $P_{b,t}$ and $P_{s,t}$ are known at time slot $t$. To avoid energy arbitrage,
the buying price is strictly greater than the selling price at any time, \ie\  $P_{b,t}>P_{s,t}$. Let $E_t$ denote the amount of energy bought from the CG at time slot $t$. Let $Q_t$ denote the portion of $E_t$ that is stored into the battery. The remaining portion $E_t-Q_t$  directly  serves the customer load. Let $F_{s,t}$ denote the amount of energy  from the battery that is sold back to the CG. The total energy sold  from the battery and the RG is bounded by
\begin{equation}
\label{equ:Et_constraint}
F_{s,t}+S_{s,t} \in[0,U_{\max}]
\end{equation}
where $U_{\max}$ is the maximum amount of energy that can be sold back to the CG\footnote{This amount may be regulated by the utility.}.

Note that while  $S_{s,t}$ from the RG can be sold back to the CG at any time, energy buying from or selling  to the CG should not happen at the same time to avoid energy arbitrage, which is ensured by the constraint $P_{b,t}>P_{s,t}$. With this constraint, we expect the following condition to be satisfied
\begin{equation}
\label{equ:sell buy constraint}
E_t\cdot F_{s,t}=0.
\end{equation}
We will verify that our proposed algorithm satisfies \eqref{equ:sell buy constraint} in Section~\ref{sec:PA}.

\subsubsection{Battery Storage} The battery operation for storage  causes the battery to deteriorate, contributing to the storage cost that has been ignored in many prior works. We model battery charging and discharging activities and the degradation cost associate to them as follows.

i) {\bf Storage operation}: The battery can be charged from multiple sources (\ie the CG and the RG) at the same time.
The total charging amount at time slot $t$ should satisfy
\begin{align}
S_{c,t}+Q_t\in[0,R_{\max}]\label{eqn:rc_bds}
\end{align}
where $R_{\max}$ is the maximum charging amount per time slot. Similarly, energy stored in the battery can be used to either serve the customer load and/or sell back to the CG. Let $F_{d,t}$ denote the amount of energy from the battery used to serve the customer at time slot $t$.
The total amount of energy discharged is bounded by
\begin{equation}
\label{eqn:dc_bds}
F_{d,t}+F_{s,t}\in[0,D_{\max}]
\end{equation}
where $D_{\max}$ is the maximum discharging amount per time slot. We assume that there is no simultaneous charging and discharging, \ie
\begin{align} \label{eqn:rc_dc_constr}
(S_{c,t}+Q_t)\cdot(F_{d,t}+F_{s,t})= 0.
\end{align}
Let $B_t$ be the battery energy level at time slot $t$, bounded by
\begin{equation} \label{eqn:Yt bds}
 B_t\in [B_{\min}, B_{\max}]
\end{equation}
where $B_{\min}$ and $B_{\max}$ are the minimum  energy required and maximum energy allowed in the battery, respectively, whose values depend on the battery type and capacity. Based on charging and discharging activities and taking the battery charging/discharging inefficiency into account, $B_t$ evolves over time as
\begin{equation}
\label{eqn:dynamics of SOB}
B_{t+1}=B_t+\eta_c(S_{c,t}+Q_t)-(F_{d,t}+F_{s,t})/\eta_d
\end{equation}
where $\eta_c$ and $\eta_d$ denote the battery charging and discharging efficiency factors, respectively, with $\eta_c,\eta_d\in[0,1]$.

Finally, the demand-and-supply balance requirement is given by
\begin{equation}
\label{eqn:Wt_constraint}
W_t=E_t-Q_t+S_{w,t}+F_{d,t}.
\end{equation}

ii) {\bf Battery degradation cost}:
It is well known that frequent charging/discharging activities cause a battery to degrade \cite{Ramadrass:JPS02}.  We model two types of battery degradation cost:  \emph{entry cost} and \emph{usage cost}. The entry cost is a fixed cost incurred due to each charging or discharging activity.  Define two indicator functions to represent charging and discharging activities: ${1_{R,t}=\{1: \textrm{if}\ Q_t+S_{c,t}>0; \; 0: \textrm{otherwise}\}}$ and ${1_{D,t}=\{1: \textrm{if}\ F_{d,t}+F_{s,t}>0; \; 0: \textrm{otherwise}\}}$. Denote the entry cost for charging by $C_{\textrm{rc}}$ and that for discharging by $C_{\textrm{dc}}$. The entry cost for battery usage at time slot $t$ is given by $x_{e,t}\triangleq 1_{R,t}C_{\textrm{rc}}+1_{D,t}C_{\textrm{dc}}$.

The battery usage cost is the cost associated with the charging/discharging amount. Define the net change of battery energy level  at time slot $t$ by  $x_{u,t}\defeq
\left|\eta_c(Q_t+S_{c,t})-(F_{d,t}+F_{s,t})/\eta_d\right|$. From \eqref{eqn:rc_bds} and \eqref{eqn:dc_bds}, it follows that $x_{u,t}$ is bounded by
\begin{align} \label{eqn:x_2}
x_{u,t} \in [0, \Gamma]
\end{align}
where $\Gamma\triangleq \max\{\eta_cR_{\max},D_{\max}/\eta_d\}$.\footnote{Accurate modeling of the battery deterioration due to charging and discharging activities is a challenging problem. Some recent work provides more detailed study on practical modeling of deterioration due to battery operation \cite{Shi&etal:ArXiv2017}.} It is known that typically faster charging/discharging, \ie larger $x_{u,t}$, has a more detrimental effect on the life time of the battery. Thus, we assume the associated cost function for usage $x_{u,t}$, denoted by $C(\cdot)$, is a continuous, convex, non-decreasing function with maximum derivative $C'(\cdot)<\infty$.

\allowdisplaybreaks

\section{ Problem Formulation}\label{sec:FHA}
For the ESM system, the system cost includes the energy buying cost minus selling profit and the battery degradation cost. Within a pre-defined $T_o$-slot time period, the average net cost of energy buying and selling  over the CG  is given by  $\overline{J}\defeq \frac{1}{T_o}\sum_{t=0}^{T_o-1}E_tP_{b,t}-(F_{s,t}+S_{s,t})P_{s,t}$. For the battery operation, the average entry cost and average net change over the $T_o$-slot period are respectively given by
\begin{align} \label{time-avg}
\overline{x_e}
&\triangleq
\frac{1}{T_o}\sum_{t=0}^{T_o-1}x_{e,t}, \quad
\overline{x_u}
\triangleq
\frac{1}{T_o}\sum_{t=0}^{T_o-1}x_{u,t}
\end{align}
where by \eqref{eqn:x_2}, $\overline{x_u}$ is bounded by
\begin{align}
\label{eqn:avg_rc_dc_bds}
\overline{x_u}\in [0, \Gamma],
\end{align}
and the battery average usage cost is $C(\overline{x_u})$. Thus, the average battery degradation cost over the $T_o$-slot period is  $\overline{x_e}+C(\overline{x_u})$.

Denote the system inputs by $\mubf_t\triangleq[W_t,S_t,P_{b,t},P_{s,t}]$ and the control actions for energy storage management by $\abf_t\triangleq[E_t,Q_t,F_{d,t},F_{s,t},S_{c,t},S_{s,t}]$ at time slot $t$. With only current input  $\mubf_t$ known, our objective is to determine a control policy $\{\pi_t\}$ for  $\abf_t$ (\ie a mapping $\pi_t:\mubf_t\rightarrow \abf_t$, $t=0,\ldots,T_o-1$.) to minimize the average system cost within the $T_o$-slot period. This is a stochastic control optimization problem and is formulated by
\begin{align}
\textrm{\bf P1:}\; &\min_{\{\pi_t\}_{t=0}^{T_o-1}} \;\;
\overline{J}+\overline{x_e}+C(\overline{x_u})\nn\\
\rm{s.t.} \;\;&
\eqref{eqn:Solar S2 S3 bounds},\eqref{equ:Et_constraint},\eqref{eqn:rc_dc_constr},\eqref{eqn:Wt_constraint},\eqref{eqn:avg_rc_dc_bds},\textnormal{and} \nonumber\\
&0\leq S_{c,t}+Q_t\leq \min\left\{R_{\max},(B_{\max}-B_t)/\eta_c\right\}\label{eqn:rc_bds_strict}\\
&0\leq F_{d,t}+F_{s,t} \leq \min\{D_{\max},\eta_d(B_t-B_{\min})\}\label{eqn:dc_bds_strict}
\end{align}
where constraints \eqref{eqn:rc_bds_strict} and \eqref{eqn:dc_bds_strict} are the derived results of  constraints \eqref{eqn:rc_bds}--\eqref{eqn:dynamics of SOB}.

{\bf  P1} is a difficult stochastic  optimization problem due to the finite time period and the correlated control actions \{$\abf_t$\} over time as a result of time-coupling dynamics of $B_t$ in  \eqref{eqn:dynamics of SOB}. Furthermore, the input distributions in\ $\mubf_t$ are unknown. The optimal control policy is difficult to obtain. Instead, we develop a real-time  algorithm for  control action $\abf_t$ over time which provides a suboptimal solution for {\bf P1} with  the cost objective being reduced as small as possible. In the following, we fist summarize our approach and technique to develop this real-time algorithm and then present details.

\subsection{Summary of Our Approach}
We develop  our real-time algorithm for {\bf P1} using    the technique of Lyapunov optimization \cite{book:Neely}, which is powerful to design dynamic control. However, to use Lyapunov optimization, the problem needs to have time-averaged objective function and constraints, which is not the case for {\bf P1}. To overcome these difficulties, we first modify {\bf P1} and then transfer the modified problem to an equivalent problem {\bf P3} that belongs to the class of stochastic optimization problem that Lyapunov optimization technique can be applied to. Then, using Lyapunov technique, we develop our real-time algorithm to determine control action $\abf_t$ for {\bf P3}. Our algorithm is to solve a per-slot opportunistic optimization problem {\bf P4} for $\abf_t$ at each time slot $t$, for which the  solution is presented in Section~\ref{sec:RT alg}. Finally, since $\{\abf_t\}$ is obtained for {\bf P3}, we design parameters of our algorithm in Section~\ref{subsec:feasibility} to ensure that it is also feasible to the original problem {\bf P1}.

\subsection{Problem Modification and Transformation}
\subsubsection{Modification}
To make {\bf P1} tractable, we first  remove the coupling of control actions over time by modifying the constraints  \eqref{eqn:rc_bds_strict} and \eqref{eqn:dc_bds_strict} on the per-slot charging and discharging amounts. We set the  change of $B_t$ over the $T_o$-slot period, \ie $B_{T_o}-B_0$, to be a desired value $\Delta_a$. From \eqref{eqn:dynamics of SOB}, this means
\begin{align}
\label{eqn:delta_a}
\frac{1}{T_o}\sum_{\tau=0}^{T_o-1}\left[\eta_c(Q_\tau+S_{r,\tau})-(F_{d,\tau}+F_{s,\tau})/\eta_d\right]=\frac{\Delta_a}{T_o}
\end{align}
where by \eqref{eqn:rc_bds}\eqref{eqn:dc_bds}\eqref{eqn:Yt bds}, we have $|\Delta_a|\leq\Delta_{\max}$, with $\Delta_{\max} \triangleq \min\{B_{\max}-B_{\min}, T_o\max\{\eta_cR_{\max},D_{\max}/\eta_d\}\}$. We now modify {\bf P1} to the follow optimization problem
\begin{align}
\textrm{\bf P2:}\; &\min_{\{\pi_t\}_{t=0}^{T_o-1}} \;\;
\overline{J}+\overline{x_e}+C(\overline{x_u})\nn\\
\rm{s.t} \;\;
&\eqref{eqn:Solar S2 S3 bounds},\eqref{equ:Et_constraint},\eqref{eqn:rc_bds}, \eqref{eqn:dc_bds}, \eqref{eqn:rc_dc_constr},\eqref{eqn:Wt_constraint},\eqref{eqn:avg_rc_dc_bds},\eqref{eqn:delta_a}.\nn
\end{align}

From {\bf P1} to {\bf P2}, by imposing the new constraint \eqref{eqn:delta_a}, we remove the dependency of per-slot charging/discharging amount on $B_t$ in constraints \eqref{eqn:rc_bds_strict} and \eqref{eqn:dc_bds_strict}, and replace them by \eqref{eqn:rc_bds} and \eqref{eqn:dc_bds},  respectively.

We point out that $\Delta_a$ in \eqref{eqn:delta_a} is  only a desired value we set in {\bf P2}. For a feasible control algorithm for  {\bf P1}, only the constraints in {\bf P1} need to be satisfied. Thus, our designed control algorithm may not satisfy  constraint \eqref{eqn:delta_a}  at the end of $T_o$-slot period.\footnote{We use {\bf P2} as an intermediate step to design the control algorithm for {\bf P1}. Thus, the proposed algorithm provides an approximate solution for {\bf P2} that may not satisfy \eqref{eqn:delta_a}.} This point is further discussed in Section~\ref{subsec:feasibility}, and in Section~\ref{sec:PA}, we will quantify the amount of mismatch with respect to $\Delta_a$ under our proposed control algorithm.

\subsubsection{Problem Transformation}
The objective of {\bf P2} contains   $C(\overline{x_u})$ which is a cost function of a time-averaged net change $\overline{x_u}$. Directly dealing with such function is difficult. Adopting the technique introduced in \cite{Neely:ArXiv2010}, we transform the problem to one that contains the time-averaged  function. To do so, we introduce an auxiliary variable $\gamma_t$ and its time average  $\overline{\gamma}\defeq \frac{1}{T_o}\sum_{\tau=0}^{T_o-1}\gamma_t$ satisfying
\begin{align}
&\gamma_t \in [0, \Gamma], \ \forall t\label{eqn:gamma_bds}\\
&\overline{\gamma}=\overline{x_u}. \label{avg_r=avg_x}
\end{align}
These constraints ensure that the auxiliary variable $\gamma_t$ and $x_{u,t}$ lie in the same range and have the same time-averaged behavior. Define $\overline{C(\gamma)}\triangleq \frac{1}{T_o}\sum_{t=0}^{T_o-1}C(\gamma_t)$
as the time average of $C(\gamma_t)$.  By using $\gamma_t$ instead of $x_{u,t}$, we replace constraint \eqref{eqn:avg_rc_dc_bds} with \eqref{eqn:gamma_bds} and \eqref{avg_r=avg_x}, and transform  {\bf P2} into the following  problem which is to optimize the control policy $\{\pi_t'\}$ for $(\gamma_t,\abf_t)$ (\ie $\pi_t':\mubf_t\rightarrow (\gamma_t,\abf_t)$, $t=0,\ldots,T_o-1$) to minimize the $T_o$-slot time average of system cost
\begin{align}
\textrm{\bf P3:}\; &\min_{\{\pi_t'\}_{t=0}^{T_o-1}} \;\;
\overline{J}+\overline{x_e}+\overline{C(\gamma)}\nn\\
\rm{s.t} \;\;
&\eqref{eqn:Solar S2 S3 bounds},\eqref{equ:Et_constraint},\eqref{eqn:rc_bds}, \eqref{eqn:dc_bds}, \eqref{eqn:rc_dc_constr},\eqref{eqn:Wt_constraint},\eqref{eqn:delta_a},\eqref{eqn:gamma_bds},\eqref{avg_r=avg_x}. \nonumber
\end{align}

It can be shown that {\bf P2} and {\bf P3} are equivalent problems (see Appendix~\ref{appA}).
The modification and transformation from {\bf P1} to {\bf P3} has enabled us to utilize Lyapunov optimization techniques  \cite{book:Neely} to design real-time control policy to solve {\bf P3}.

\subsection{Real-Time Control via Lyapunov Optimization}

To design a real-time algorithm, in Lyapunov optimization, virtual queue is introduced for each time-averaged constraint, and Lyapunov drift for the queues is defined. It is shown that keeping the stability of each queue is equivalent to satisfying the constraint \cite{book:Neely}. Thus, instead of the original cost objective in the optimization problem, a drift-plus-cost metric is considered  in Lyapunov optimization and real-time algorithm is developed to minimize this metric. In the following, we develop our algorithm using this technique.

\subsubsection{Virtual queues}
Based on the Lyapunov framework, we introduce two virtual queues $Z_t$ and $H_t$ for time-averaged constraints \eqref{eqn:delta_a} and \eqref{avg_r=avg_x}, respectively, as
\begin{align}
Z_{t+1}&=Z_t+\eta_c(Q_t+S_{c,t})-(F_{d,t}+F_{s,t})/\eta_d-\frac{\Delta_a}{T_o}, \label{eqn:queue Z}\\
H_{t+1}&=H_t+\gamma_t-x_{u,t}. \label{eqn:queue H}
\end{align}
From \eqref{eqn:dynamics of SOB} and \eqref{eqn:queue Z}, $Z_t$ and $B_t$ have the following relation
\begin{align}\label{eqn:dynamic shift Z}
Z_t=B_t-A_t
\end{align}
where $A_t\defeq A_o+\frac{\Delta_a}{T_o}t$ in which  $A_o$ is a constant shift and $\frac{\Delta_a}{T_o}t$  ensures that the left hand side equality in \eqref{eqn:delta_a} is satisfied.
We will revisit the value of $A_o$ to  ensure a feasible solution for {\bf P1}.

\subsubsection{Lyapunov drift}
Define $\Thetabf_t\triangleq[Z_t, H_t]$. Define the quadratic Lyapunov function  $L(\Thetabf_t)\triangleq\frac{1}{2}(Z_t^2+H_t^2)$. Divide $T_o$ slots into $M$ sub-frames of $T$-slot duration as $T_o=MT$, for  $M,T\in \mathbb{N}^+$. We define a one-slot sample path Lyapunov drift as $\Delta(\Thetabf_t)\triangleq L\left(\Thetabf_{t+1}\right)-L(\Thetabf_{t})$, which only depends on the current system inputs $\mubf_t$.

\subsubsection{Drift-plus-cost metric}
We define a drift-plus-cost metric which is a weighted sum of the drift $\Delta(\Thetabf_t)$ and the system cost  at current time slot $t$, given by
\begin{align}\label{drift_cost_metric}
\hspace{-.5em}\Delta(\Thetabf_t)+V[E_tP_{b,t}-(F_{s,t}+S_{s,t})P_{s,t}+x_{e,t}+C(\gamma_t)]
\end{align}
where constant $V>0$ sets the relative weight between the drift and the system cost.In Lyapunov optimization, instead of minimizing the system cost objective in {\bf P3},  we aim to minimize this drift-to-cost metric. However, directly using this metric to design a control policy is still difficult.
Instead, we obtain an upper bound on the drift $\Delta(\Thetabf_t)$, which will be used for designing our  real-time control algorithm.
\begin{lemma}\label{lemma drift upper bound}
Lyapunov drift $\Delta(\Thetabf_t)$ is upper bounded by
\begin{align}\label{eqn:drift delta 1}
\Delta(\Thetabf_t)\leq&\
Z_t\left(E_t+S_{c,t}+S_{w,t}-W_t-F_{s,t}-\frac{\Delta_a}{T_o}\right)\nn\\
&+H_t\gamma_t-g(H_t)(E_t+S_{c,t}+l_t)+G
\end{align}
where $G \triangleq\frac{1}{2}\max\left\{(\eta_cR_{\max}-\frac{\Delta_a}{T_o})^2, (D_{\max}/\eta_d+\frac{\Delta_a}{T_o})^2\right\} +\frac{1}{2}\Gamma^2$, and
\begin{align}
g(H_t) &\triangleq \begin{cases}\eta_cH_t & H_t \ge 0 \\ H_t/\eta_d & H_t <0 \end{cases} \label{g(Ht)}\\
l_t & \triangleq \text{sgn}(H_t)(S_{w,t}-W_t-F_{s,t})
\end{align}
where $\text{sgn}(\cdot)$ is the sign function.
\end{lemma}
\IEEEproof
See Appendix \ref{app drift upper bound}.
\endIEEEproof

By Lemma~\ref{lemma drift upper bound}, an upper bound on the per-slot drift-plus-cost metric in \eqref{drift_cost_metric} is readily obtained. In the following, we use this upper bound to develop a real-time control algorithm.

\section{Real-Time Bidirectional  Control Algorithm}\label{sec:RT alg}
We now propose our real-time control algorithm  to minimize the upper bound on the drift-plus-cost metric \emph{per slot}.
Removing all the constant terms in the upper bound of the drift-plus-cost metric that are independent of $\abf_t$ and $\gamma_t$, we have the equivalent optimization problem which can be further separated into two sub problems for $\gamma_t$ and $\abf_t$, respectively, as follows
\begin{align}
 \text{\bf P4}_a:\; &\min_{\gamma_t}
\quad H_t\gamma_t+VC(\gamma_t)
\quad \textrm{s.t.} \;\; \eqref{eqn:gamma_bds}.\nn\\
\text{\bf P4}_b:\; &\min_{\abf_t}
  \quad
E_t(Z_t-g(H_t)+VP_{b,t})+S_{c,t}(Z_t-g(H_t))\nn\\
  &\quad -F_{s,t}(Z_t-|g(H_t)|+VP_{s,t})-S_{s,t}VP_{s,t}+Vx_{e,t} \nn\\
  &\textrm{s.t.} \quad
    \eqref{eqn:Solar S2 S3 bounds},\eqref{equ:Et_constraint},\eqref{eqn:rc_bds},\eqref{eqn:dc_bds},\eqref{eqn:rc_dc_constr},\eqref{eqn:Wt_constraint},\eqref{eqn:delta_a}.\nn
\end{align}

First, we solve $\text{\bf P4}_a$ to obtain the optimal solution $\gamma_t^*$.
Note that $\text{\bf P4}_{a}$ is convex for $C(\cdot)$ being convex. Thus, we can directly solve it and  obtain the optimal  $\gamma^*_t$  of $\text{\bf P4}_a$.
\begin{lemma}\label{lemma33}
The optimal solution $\gamma^*_t$ of $\text{\bf P4}_a$ is given by
\begin{align}
\label{eqn:optimal gamma}
\gamma^*_t=
    \begin{cases}
        0&\text{if}\ H_t\geq 0\\
        \Gamma& \text{if}\ H_t< -VC'(\Gamma)\\
        C'^{-1}\left(-\frac{H_t}{V}\right)&\text{otherwise}
    \end{cases}
\end{align}
where $C'(\cdot)$ is the first derivative of $C(\cdot)$, and $C'^{-1}(\cdot)$  the inverse function of $C'(\cdot)$.
\end{lemma}
\IEEEproof
See Appendix~\ref{app:lemma33}.
\endIEEEproof

Next, we obtain the optimal $\abf_t^*$ of $\text{\bf P4}_{b}$ and provide the conditions under which  $\abf_t^*$ is feasible to {\bf P1}.

\subsection{The Optimal Control $\abf^*_t$  for $\text{\bf P4}_{b}$}
Denote the objective function of $\text{\bf P4}_{b}$ by $J(\abf_t)$.
Define the \emph{idle state} of the battery as the state where there is no charging or discharging activity. The control decision in the idle state is given by $\abf_t^\textrm{id}=[E_t^\textrm{id}, Q_t^\textrm{id},F_{d,t}^\textrm{id},F_{s,t}^\textrm{id},S_{c,t}^\textrm{id},S_{s,t}^\textrm{id}]$, where $E_t^\textrm{id}=W_t-S_{w,t}$, $Q_t^\textrm{id}=F_{d,t}^\textrm{id}=F_{s,t}^\textrm{id}=S_{c,t}^\textrm{id}=0$, and $S_{s,t}^\textrm{id}=\min\{S_t-S_{w,t},U_{\max}\}$. Then, in the idle state, we have
$J(\abf_t^\textrm{id})= (W_t-S_{w,t})(Z_t-g(H_t)+VP_{b,t})-VP_{s,t}\min\{S_t-S_{w,t},U_{\max}\}$.
We derive the optimal control decision $\abf_t^*=[E_t^*,Q_t^*,F_{d,t}^*,F_{s,t}^*,S_{c,t}^*,S_{s,t}^*]$   in five cases in Proposition~\ref{prop0} below.
Define $[a]^+\triangleq \max\{a,0\}$.
\begin{proposition}\label{prop0}
 Define $(S_{c,t}^\text{a},S_{s,t}^\text{a})$  as follows:
 If $VP_{s,t}\ge {g(H_t)-Z_t}$: $S^\text{a}_{s,t}\triangleq \min\{S_t-S_{w,t},U_{\max}\}$,  $S^\text{a}_{c,t}\triangleq\min\{S_t-S_{w,t}-S^\text{a}_{s,t},R_{\max}\}$;  Otherwise, $S^\text{a}_{c,t}\triangleq\min\{S_t-S_{w,t},R_{\max}\}$,  $S^\text{a}_{s,t}\triangleq\min\{S_t-S_{w,t}-S^\text{a}_{c,t},U_{\max}\}$.

Denote $\abf_t^\w=[E_t^\w,Q_t^\w,F_{d,t}^\w,F_{s,t}^\w,S_{c,t}^\w,S_{s,t}^\w]$ as the control decision in the charging or discharging state.  The optimal control solution $\abf^*_t$ for $\bf \text{\bf P4}_b$ is given in the following cases:\newcounter{qcounter}
\begin{list}{{\it \arabic{qcounter}$\left.\right)$~}}
{\usecounter{qcounter}
\setlength\leftmargin{1.5em}
\setlength\labelwidth{2em}
\setlength\labelsep{0em}
\setlength\itemsep{.5em}
}

\item {\it For $Z_t-g(H_t)+VP_{b,t}\leq 0$}: The battery is in either the charging state or the idle state. Let
    \begin{align}\label{case a}
        \hspace*{-0.5em}\begin{cases}
        F^\w_{d,t}=F^\w_{s,t}=0,\ S^\w_{c,t}=S^\text{a}_{c,t},\ S^\w_{s,t}=S^\text{a}_{s,t}\\
        Q^\w_t=R_{\max}-S^\w_{c,t} \\
        E^\w_t=W_t-S_{w,t}+R_{\max}-S^\w_{c,t}.
        \end{cases}
  \end{align}
Then, $\abf^*_t=\arg\min_{\abf_t\in\{\abf_t^\w,\abf_t^\textrm{id}\}} J(\abf_t)$.

\item{\it For $\max\{Z_t-g(H_t),Z_t-|g(H_t)|+VP_{s,t}\}<0\leq Z_t-g(H_t)+VP_{b,t}$}: The battery is in either the charging state (from the RG only), the discharging state (to the customer load only), or the idle state. Let
  \begin{align}\label{case c}
      \begin{cases}
      F^\w_{d,t}=\min\{W_t-S_{w,t},D_{\max}\}\\
      F^\w_{s,t}=Q^\w_t=0,\ S^\w_{c,t}=S^\text{a}_{c,t},\ S^\w_{s,t}=S^\text{a}_{s,t}\\
      E^\w_t=[W_t-S_{w,t}-D_{\max}]^+.
      \end{cases}
  \end{align}
Then, $\abf^*_t=\arg\min_{\abf_t\in\{\abf_t^\w,\abf_t^\textrm{id}\}} J(\abf_t)$.

\item{\it For $Z_t-g(H_t)\le 0\le Z_t-|g(H_t)|+VP_{s,t}$}: The battery is  in either the charging state (from the RG only), the discharging state, or the idle state. Define  $\abf_t^\text{dc}$ in the discharging state as
  \begin{align}\label{case d 2}
\hspace*{-.5em}
      \begin{cases}
      F^\text{dc}_{d,t}=\min\{W_t-S_{w,t},D_{\max}\}\\
      S^\text{dc}_{s,t}=\min\{S_t-S_{w,t},U_{\max}\} \\
      F^\text{dc}_{s,t}=\min\{D_{\max}-F^\text{dc}_{d,t},U_{\max}-S^\text{dc}_{s,t}\}\\
      S^\text{dc}_{c,t}=Q^\text{dc}_t=0, \ E^\text{dc}_t=[W_t-S_{w,t}-D_{\max}]^+;
      \end{cases}
  \end{align}
Define $\abf_t^\text{rc}$ in the charging state as
  \begin{align}\label{case d 3}
\hspace*{-.5em}
      \begin{cases}
      F^\text{rc}_{d,t}=F^\text{rc}_{s,t}=Q^\text{rc}_t=0 \ \\
      S^\text{rc}_{c,t}=S^\text{a}_{c,t}, \ S^\text{rc}_{s,t}=S^\text{a}_{s,t}\\
      E^\text{rc}_t= W_t-S_{w,t}.
      \end{cases}
  \end{align}
Then, $\abf^*_t=\arg\min_{\abf_t\in\{\abf_t^\text{rc},\abf_t^\text{dc},\abf_t^\textrm{id}\}} J(\abf_t)$.

\item{\it For $Z_t-|g(H_t)|+VP_{s,t}<0 \le Z_t-g(H_t)$}: The battery is in either the discharging state (to the customer load only) or the idle state. Let
  \begin{align}\label{case f}
      \begin{cases}
      F^\w_{d,t}=\min\{W_t-S_{w,t},\ D_{\max}\}\\  F^\w_{s,t}=Q^\w_t=0\\
      S^\w_{s,t}=\min\{S_t-S_{w,t},U_{\max}\},S^\w_{c,t}=0\\
      E^\w_t=[W_t-S_{w,t}-D_{\max}]^+.
      \end{cases}
  \end{align}
Then, $\abf^*_t=\arg\min_{\abf_t\in\{\abf_t^\w,\abf_t^\textrm{id}\}} J(\abf_t)$.

\item{\it For $\min\{Z_t-g(H_t),Z_t-|g(H_t)|+VP_{s,t}\}>0$}: The battery is in either the discharging state or the idle state. If $Z_t>|g(H_t)|$, let
  \begin{align} \label{case b 1}
      \begin{cases}
      F^\w_{d,t}=\min\{W_t-S_{w,t},D_{\max}\}\\
      F^\w_{s,t}=\min\{D_{\max}-F^\w_{d,t},U_{\max}\}\\
      S^\w_{s,t}=\min\{S_t-S_{w,t},U_{\max}-F^\w_{s,t}\} \\
      S^\w_{c,t}= Q^\w_t=0,\      E^\w_t=[W_t-S_{w,t}-D_{\max}]^+;
      \end{cases}
  \end{align}
Otherwise,
let  \begin{align} \label{case b 2}
      \begin{cases}
      F^\w_{d,t}=\min\{W_t-S_{w,t},D_{\max}\}\\
      S^\w_{s,t}=\min\{S_t-S_{w,t},U_{\max}\} \\
      F^\w_{s,t}=\min\{D_{\max}-F^\w_{d,t},U_{\max}-S^\w_{s,t}\}\\
      S^\w_{c,t}= Q^\w_t=0, \ E^\w_t=[W_t-S_{w,t}-D_{\max}]^+.
      \end{cases}
  \end{align}
Then, $\abf^*_t=\arg\min_{\abf_t\in\{\abf_t^\w,\abf_t^\textrm{id}\}} J(\abf_t)$.

\end{list}
\end{proposition}
\IEEEproof
See Appendix~\ref{app:p4b}.
\endIEEEproof

Proposition ~\ref{prop0} provides the closed-form control solution in five cases, depending on the battery energy level (via $Z_t$), battery usage cost (via $H_t$), and the prices. In each case, $J(\abf^\w_t)$ in the charging (or discharging) state is compared with $J(\abf_t^\textrm{id})$ in the idle state, and the optimal   $\abf_t^*$ is the control solution of the state with the minimum objective value.

\emph{Remarks:} Note that there are two sources to be controlled for  selling energy back: $F_{s,t}$ from the battery and $S_{s,t}$\ from the RG. From Proposition~\ref{prop0}, whether or not to sell energy from the battery back to the grid depends on the battery energy level.  When the battery energy level is low (Case 1), energy is kept in the battery. When the battery has a moderately low energy level (Case 2), it may be  in either the charging or discharging state. For the latter,  the battery only supplies enough energy to the customer but does not sell energy back. When the battery  energy level is higher but still moderate (Case 3),  it may still be  in either the charging or discharging state. For the latter, the battery may sell energy back to the grid. When the battery has just sufficient energy (Case 4), it may supply energy  to the customer, but will not sell energy back to the grid.  When the energy level in the battery is high (Cases 5), it may  supply energy to the customer and at the same time sell energy back. In contrast,  the renewable energy can be sold to the grid regardless of the battery energy level,  state (charging, discharging, or idle), or the price  to make an additional profit. As the result, energy generated by the renewable will be utilized as much as possible. However, when  the system  wants to sell energy from both the battery and the renewable, the order to determine $S_{s,t}$ and $F_{s,t}$ depends on which results in the minimum cost in $\text{\bf P4}_b$. In Case 5, for the control decision in \eqref{case b 1},   $S_{s,t}$ is determined after $F_{s,t}$, while in \eqref{case b 2},  $F_{s,t}$ is determined after $S_{s,t}$.

\subsection{Feasible $\{\abf_t^*\}$ for {\bf P1}}\label{subsec:feasibility}
The optimal solution  $\mathbf{a}^*_t$ of $\text{\bf P4}_b$ provides a real-time solution for {\bf P3}. However, it may not be feasible to {\bf P1}, because  the battery  capacity constraint \eqref{eqn:Yt bds} on $B_t$ may be violated. By properly designing $A_o$ and $V$, we can guarantee that $\abf_t^*$ satisfies  constraint \eqref{eqn:Yt bds}, and ensure the feasibility of the solution. Define $[a]_-\triangleq\min\{a,0\}$. The result is stated below.
\begin{proposition} \label{prop1}
For the proposed real-time control algorithm, by setting $V\in (0,V_{\max}]$ with
\begin{align}\label{eqn:V_max}
&V_{\max}= \nn \\
&\ \ \frac{B_{\max}-B_{\min}-\eta_c R_{\max}-(D_{\max}+2\Gamma)/\eta_d-|\Delta_a|}{P_{b}^{\max}+C'(\Gamma)/\eta_d+[C'(\Gamma)/\eta_d-P_{s}^{\min}]^+},
\end{align}
and
 $A_t$ in \eqref{eqn:dynamic shift Z} with
\begin{align}\label{eqn:A_o}
 A_o = \!\! \ B_{\min}\!+\!VP_{b}^{\max}\!+\!\frac{VC'(\Gamma)\!+\Gamma\!+\!D_{\max}}{\eta_d}\!+\!\frac{\Delta_a}{T_o} \!-\![\Delta_a]_-,
\end{align}
 $B_t$ satisfies the battery capacity constraint \eqref{eqn:Yt bds}, and  control solution $\abf_t^*$ of $\text{\bf P4}_b$, for any $t$, is feasible to {\bf P1}.
\end{proposition}
\IEEEproof
See Appendix \ref{appC}.
\endIEEEproof

Note that  $V_{\max}>0$ in \eqref{eqn:V_max} is generally satisfied for practical battery storage capacity and $|\Delta_a|$ being set relatively small.
We should also point out that since $\Delta_a$ is a desired value set by our proposed algorithm, the solution $\abf_t^*$ of ${\bf P4}_b$ may not necessarily satisfy constraint \eqref{eqn:delta_a} at the end of the $T_o$-slot period, and thus may not be feasible to {\bf P2}. However, Proposition~\ref{prop1} provides the values of $A_o$ and $V$ to guarantee the control solutions $\{\abf_t^*\}$ being feasible to {\bf P1}.

The proposed real-time bidirectional energy control algorithm is summarized in Algorithm~\ref{alg1}. We emphasize that  i) our proposed algorithm does not require or rely on any statistical distributions of system inputs $\mubf_t$ (prices, load, and renewable generation processes), and thus can be applied to general scenarios, especially when these processes are non-ergodic or difficult to predict in a highly dynamic environment. ii) The control solution provided by our real-time algorithm is given in closed-form which requires little computation, and thus is attractive for practical implementation.

\begin{algorithm}[t]
\caption{Real-time battery management control algorithm}\label{alg1}
Initialize: $Z_0=H_0=0$.\\
Determine $T_o$.\\
Set $\Delta_a\in [-\Delta_{\max},\Delta_{\max}]$.\\
Set $A_o$ and $V\in(0,V_{\max}]$ as in \eqref{eqn:A_o} and \eqref{eqn:V_max}, respectively.\\
At time slot $t$:
\begin{algorithmic}[1]
\State Observe the system inputs $\mubf_t$ and queues $Z_t$ and $H_t$.
\State Solve ${\bf \text{\bf P4}_{a}}$ and obtain $\gamma^*_t$ in \eqref{eqn:optimal gamma}; Solve ${\bf \text{\bf P4}_{b}}$ and obtain $\mathbf{a}^*_t$ by following cases \eqref{case a}-\eqref{case b 2} in Proposition~\ref{prop0}.
\State Use $\mathbf{a}^*_t$ and $\gamma^*_t$ to update $Z_{t+1}$ and $H_{t+1}$ in \eqref{eqn:queue Z} and \eqref{eqn:queue H}, respectively.
\State Output control decision $\abf_t^*$.
\end{algorithmic}
\end{algorithm}

\section{Performance Analysis}\label{sec:PA}
To analyze the performance of our real-time solution in Algorithm~\ref{alg1} with respect to {\bf P1}, let $u^*(V)$ denote the $T_o$-slot average system cost objective  of {\bf P1} achieved by Algorithm~\ref{alg1}, which depends on  the value of $V$ set by Algorithm~\ref{alg1}.   For comparison, we partition $T_o$ slots into $T$ frames with $T_o=MT$, for some integers $M,T\in \mathbb{N}^+$. Within each frame $m$, we consider a \emph{$T$-slot look-ahead optimal control policy}, where $\{W_t,S_t,P_{b,t},P_{s,t}\}$ are known non-causally for the entire frame beforehand. Let $u_m^\textrm{opt}$ denote the minimum $T$-slot average cost for frame $m$ achieved by this optimal policy. We can view $u_m^\textrm{opt}$ as the minimum objective value of {\bf P1} with $T_o = T$ under the optimal $T$-slot look-ahead solution. The  performance gap of our proposed real-time algorithm  to the optimal $T$-slot look-ahead policy is bounded in the following theorem.

\begin{theorem} \label{thm1}
For any arbitrary system inputs $\{\mubf_t\}$, and any $M,T\in \mathbb{N}^+$ with $T_o=MT$, the  $T_o$-slot average system cost under  Algorithm~\ref{alg1}  to that under the optimal $T$-slot look-ahead policy satisfies
\begin{align}\label{thm1:bd}
&u^*(V)-\frac{1}{M}\sum_{m=0}^{M-1}u_m^\textrm{opt} \nn\\
&\leq
\frac{G T}{V}+\frac{L(\Thetabf_0)-L(\Thetabf_{T_o})}{VT_o}+\frac{C'(\Gamma)(H_0-H_{T_o})}{T_o}
\end{align}
with the bound at the right hand side being finite.
Asymptotically as $T_o\to \infty$, \\[-1.5em]
\begin{align}\label{thm1:bd_longterm}
\lim_{T_o\to \infty}u^*(V)-\lim_{T_o\to \infty}\frac{1}{M}\sum_{m=0}^{M-1}u_m^\textrm{opt}\le \frac{G T}{V}.
\end{align}
\end{theorem}
\IEEEproof
See Appendix~\ref{appD}.
\endIEEEproof

By Theorem~\ref{thm1}, the performance gap of Algorithm~\ref{alg1} to the $T$-slot look-ahead optimal policy is upper bounded in \eqref{thm1:bd}, for any $T$ with $T_o=MT$. To minimize the gap, we should always set $V = V_{\max}$. From \eqref{thm1:bd_longterm}, as the duration goes to infinity, the asymptotic gap is in the order of $\Oc(1/V)$. Since $V_{\max}$ increases with $B_{\max}$, When $V=V_{\max}$, Algorithm~\ref{alg1} is asymptotically equivalent to the $T$-slot look-ahead optimal policy as the battery capacity and time duration increases.

As discussed at the end of Section~\ref{subsec:feasibility}, constraint \eqref{eqn:delta_a} in {\bf P2} sets a desired value for  $\Delta_a$ which may not be achieved by our proposed algorithm at the end of $T_o$ slots. Denote this mismatch under Algorithm~\ref{alg1} by $\epsilon\triangleq \sum_{\tau=0}^{T_o-1}(Q_\tau+S_{r,\tau}-F_{s,\tau}-F_{d,\tau})- \Delta_a$. This mismatch is quantified below.
\begin{proposition}\label{prop3}
For any arbitrary system inputs $\{\mubf_t\}$ and  any initial queue value $Z_{0}\in \mathbb{R}$, the mismatch for constraint \eqref{eqn:delta_a} by Algorithm~\ref{alg1} is given by $\epsilon = Z_{T_o}-Z_{0}$, and is bounded by
\begin{align}
|\epsilon|\le & \  2\Gamma/\eta_d+VC'(\Gamma)/\eta_d+V[ C'(\Gamma)/\eta_d-P_{s}^{\min}]^+\nn \\
&+VP_{b}^{\max}+\eta_cR_{\max}+D_{\max}/\eta_d.
\end{align}
\end{proposition}
\IEEEproof
See Appendix~\ref{appE}.
\endIEEEproof

Finally, we expect constraint \eqref{equ:sell buy constraint} to be satisfied by Algorithm~\ref{alg1}, \ie buying energy ($E_t>0$) and selling stored energy  ($F_{s,t}>0$) should not occur at same time. This is verified in the following result.
\begin{proposition}\label{prop E times P zero}
For any system inputs $\mubf_t$, the optimal control solution $\abf^*_t$ under Algorithm~\ref{alg1} guarantees constraint \eqref{equ:sell buy constraint} being satisfied.
\end{proposition}
\IEEEproof
See Appendix~\ref{app E times P zero}.
\endIEEEproof

\section{Simulation Results}\label{sec:sim}

We set the slot duration to be 5 minutes, and assume that system input  $\mubf_t$ remains unchanged within each slot. We set the buying price $P_{b,t}$ using the data collected from Ontario Energy Board \cite{website:Ontario_Elec}. We use solar energy for the RG to generate  $\{S_t\}$. As a result,  $\{S_t\}$ is a non-ergodic process, with the mean  $\overline{S}_t=\mathbb{E}[S_t]$ changing periodically over $24$ hours. As shown in Fig.~\ref{fig:C_S_W_distribution} top, we model $\{\overline{S}_t\}$ by  a three-stage pattern as $\{\overline{S^{h}},\overline{S^{m}},\overline{S^{l}}\}=\{1.98,0.96,0.005\}/12$~{kWh,} and set standard deviation $\sigma_{S^i}=0.4\overline{S^i}$, for $i=h,m,l$. We also model the load $W_t$ as a non-ergodic process with mean  $\overline{W}_{t}=\mathbb{E}[W_t]$ following a three-stage  pattern over each day as $\{\overline{W^{h}},\overline{W^{m}},\overline{W^{l}}\}=\{2.4,1.38,0.6\}/12$~{kWh}, shown in Fig.~\ref{fig:C_S_W_distribution} middle, and set standard deviation  $\sigma_{W^i}=0.2\overline{W^i}$, for $i=h,m,l$. Buying price $P_{b,t}$ follows a three-stage price pattern repeated each day as $\{P_{b}^h, P_{b}^m, P_{b}^l\}=\{\$0.118, \$0.099, \$0.063\}$ per kWh, as shown in Fig.~\ref{fig:C_S_W_distribution} bottom. The battery and storage parameters are set as follows: $B_{\min}=0$, $R_{\max}=0.15$~{kWh}, $D_{\max}=0.15$~{kWh}, $C_{\textrm{rc}}=C_{\textrm{dc}}=0.001$, $U_{\max}=0.2$~{kWh},  and the initial battery energy level $B_0={B_{\max}}/{2}$. Unless specified, we set the following default values: $B_{\max}=3$~kWh, and $\eta_c=\eta_d=0.98$.\footnote{A typical Lithium-ion battery can achieve $0.99$ charging efficiency.}

\emph{Quadratic battery usage cost}:
We use a quadratic function for the battery usage cost as an exemplary case, given by $C(\overline{x_u})=k\overline{x_u}^2$, where  $k>0$ is the battery cost coefficient depending on the battery specification and $\overline{x_u}$ is given in \eqref{time-avg}. The optimal $\gamma_t^*$  of $\text{\bf P4}_a$ in \eqref{eqn:optimal gamma} in this case can be derived as:
i) $\gamma^*_t= 0$ for $H_t> 0$; ii) $\gamma^*_t=\Gamma$ for $H_t< -kV\Gamma$; iii) $\gamma^*_t=-{H_t}/(2kV)$ for $-2kV\Gamma \le H_t \le 0$.
We use this cost function throughout our simulation study. Unless specified, we set $k=0.1$ as the default value.

We consider a 24-hour duration with  $T_o=288$ slots. Since a positive (negative)   $\Delta_a$ allows battery to charge (discharge) more than discharge (charge) over a $T_o$-period,  we alternate the sign of $\Delta_a$  over multiple $T_o$-slot periods to control this tendency:  we set $\Delta_a=+c$ ($-c$) for the odd (even) $T_o$-slot periods, for some constant $c>0$. Unless specified, we set $V=V_{\max}$ as the default value.

\subsubsection{Energy buying and selling vs. prices}
\begin{figure}[t]
\centering
\includegraphics[width=0.8\linewidth]{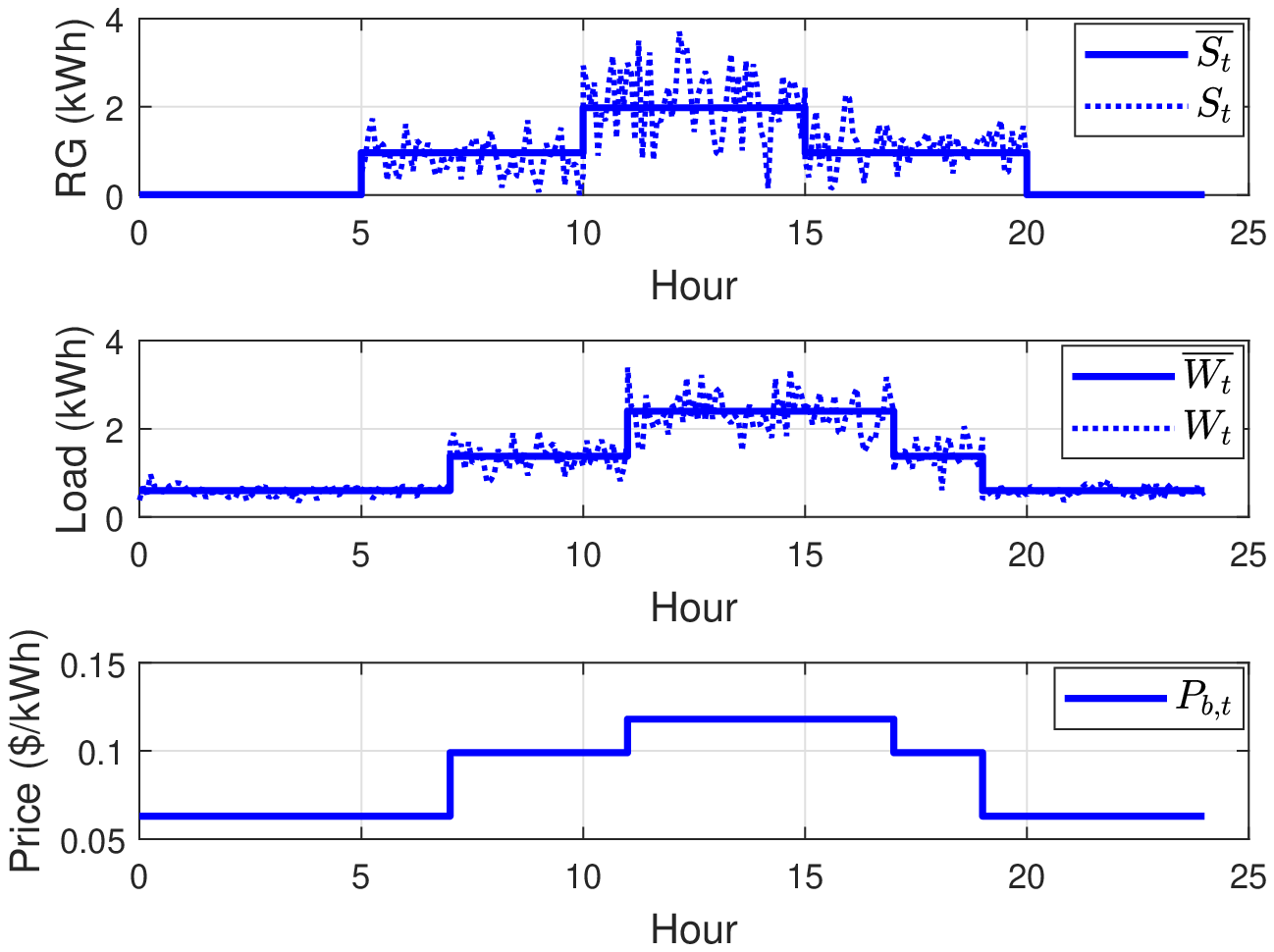}\vspace*{-1em}
\caption{\small Example  of $W_t$, $S_t$ and $P_{b,t}$ over 24 hours. }
\label{fig:C_S_W_distribution}
\centering
\includegraphics[width=0.8\linewidth]{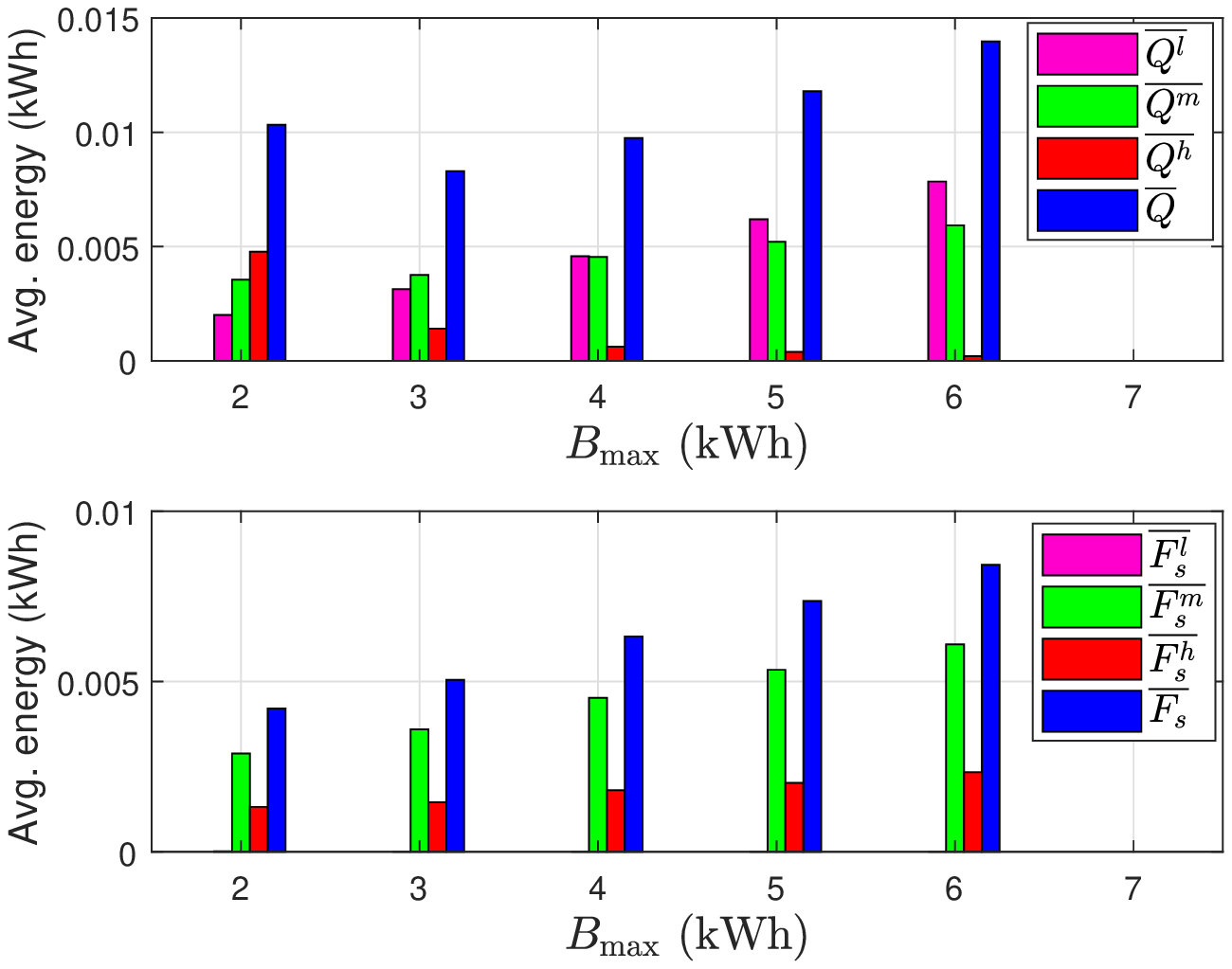}
\caption{\small Average energy bought (sold) into (from) the battery at different prices  (selling-to-buying price ratio $\rho_p=0.9$).}
\label{fig:bar_90per}
\centering
\includegraphics[width=0.8\linewidth]{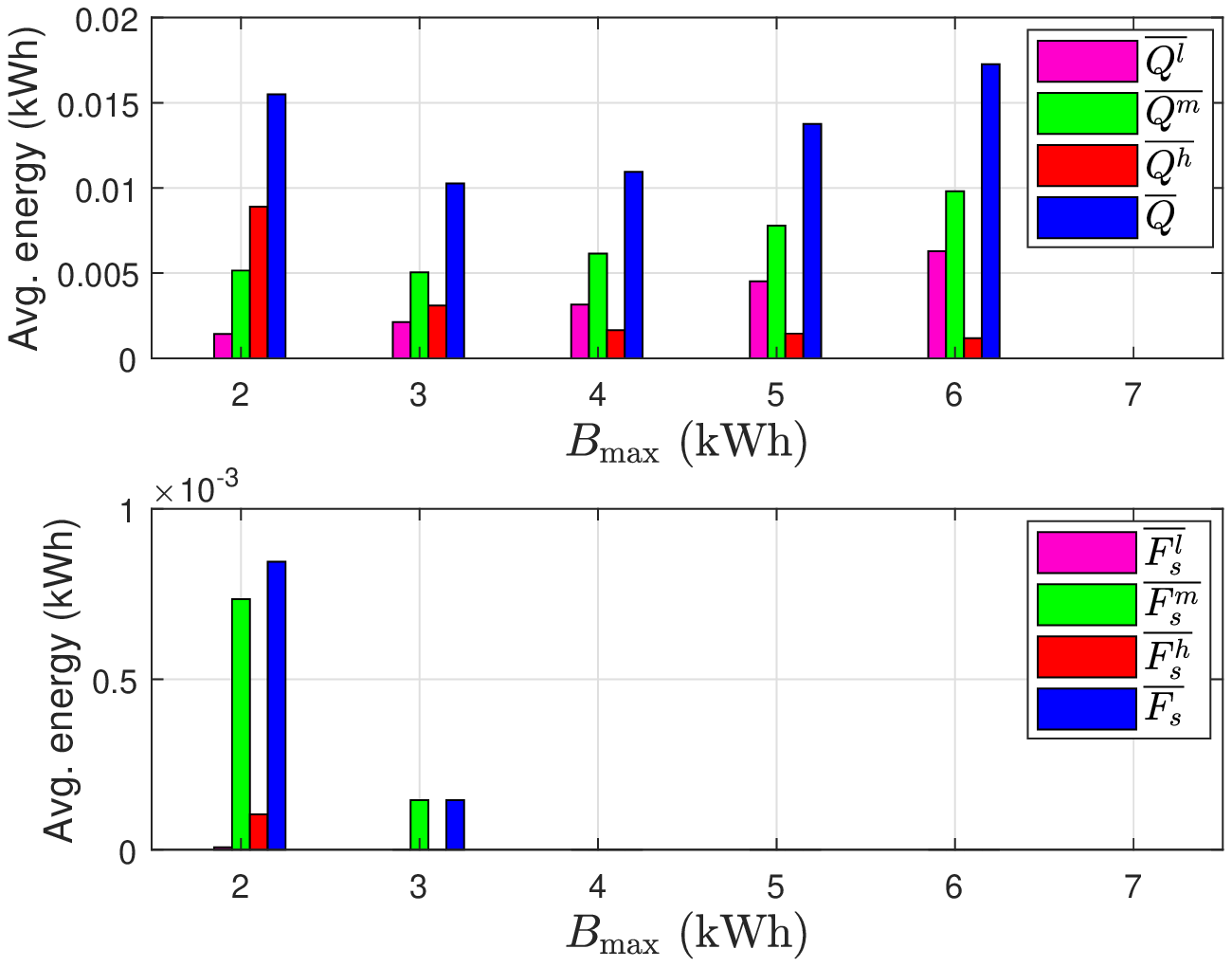}
\caption{\small  Average energy bought (sold) into (from) the battery at different prices  (selling-to-buying price ratio $\rho_p=0.3$).}
\label{fig:bar_30per}\vspace*{-1em}
\end{figure}
We  study the energy  buying and selling behaviors under different selling-to-buying price ratios. Define the selling-to-buying price ratio by $\rho_p\triangleq {P_{s,t}}/{P_{b,t}}$ with $0\le \rho_p<1$. It is fixed over $T_o$ time slots. Define the average amount of bought and sold energy at each price stage of $P_{b,t}$  (high, medium, low)  as
\begin{align*}
\hspace*{-1em}\overline{Q^i}\triangleq \frac{1}{|\Tc_b^i|}\sum_{t\in \Tc_b^i}Q_t^*, \
\overline{F_s^i}\triangleq \frac{1}{|\Tc_s^i|}\sum_{t\in \Tc_s^i}F_{s,t}^*, \ i=h,m,l
\end{align*}
where $\Tc_b^i\triangleq\{t:P_{b,t}=P_b^i\}$ and $\Tc_s^i\triangleq\{t:P_{s,t}= P_s^i\}$. Also, denote $\overline{Q}$ and $\overline{F_s}$ as the overall energy bought and sold, respectively.  Figs.~\ref{fig:bar_90per} and \ref{fig:bar_30per} show the average amount of energy for $\rho_p=0.9$ and $0.3$ (high and low selling prices), respectively.
Comparing Fig.~\ref{fig:bar_90per} bottom with Fig.~\ref{fig:bar_30per} bottom, we see that more energy is sold  at higher $\rho_p$ to increase the profit, while at lower $\rho_p$, selling at the price is not cost-effective, and the system tends to keep the energy in the storage  for future usage.
Furthermore, in Fig.~\ref{fig:bar_90per}  bottom,  at higher $\rho_p$, the average amount of energy sold  at $P_s^m$ and $P_s^h$ increases with  the battery capacity $B_{\max}$. This is because a larger capacity offers more flexibility for charging and discharging activities, and allows more energy to be sold back at higher prices. For the same reason, a larger capacity allows more energy to be bought at lower price, as shown in  Figs.~\ref{fig:bar_90per} and \ref{fig:bar_30per} top, where, as $B_{\max}$ increases, the amount of energy bought from the grid at higher price $P_{b,t}=P_b^h$ decreases and at lower price $P_{b,t}=P_b^l$ increases.
\subsubsection{Setting $\Delta_a$ and mismatch $\epsilon$}
\begin{figure}[t]
\centering
\includegraphics[width=0.8\linewidth]{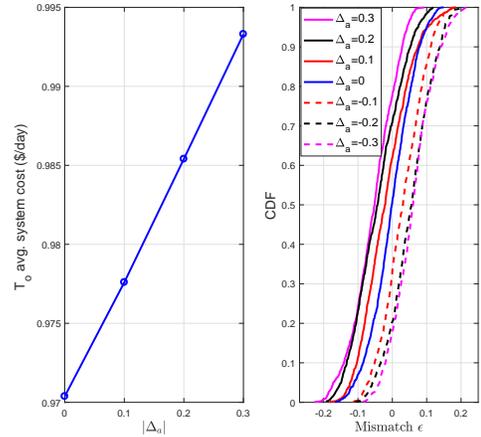}
\caption{\small  Performance of Algorithm~\ref{alg1}  at different  $\Delta_a$ setting (selling-to-buying price ratio $\rho_p=0.9$). Left: average system cost at different $|\Delta_a|$; Right: CDF of mismatch $\epsilon$ at different $\Delta_a$. }
\label{fig:delta}
\end{figure}
Fig.~\ref{fig:delta} shows how the average system cost  varies with  $\Delta_a$ set by proposed Algorithm~\ref{alg1}, for $\rho_p=0.9$. We simulate the system over multiple  $T_o$-slot periods for a given $|\Delta_a|$. It shows that the system cost increases with $|\Delta_a|$. More energy needs to be bought to be stored in the battery for $\Delta_a>0$,  leading to a higher system cost, while it is the opposite for $\Delta_a<0$. Overall, the system cost is the smallest when $\Delta_a=0$.  Fig.~\ref{fig:delta} right shows the CDF of the mismatch $\epsilon$ at different value of   $\Delta_a$ by Algorithm~\ref{alg1}. We see that $\epsilon$ at different values of $\Delta_a$ is relatively small for $\rho_p=0.9$, as the availability to sell energy helps keep the mismatch  relatively small. Also from the sign of $\epsilon$, it shows that the net energy change in the battery over the period tends to be less than $|\Delta_a|$.  The mismatch $\epsilon$ is the smallest when $\Delta_a=0$. Based on the above results, we conclude that  setting $\Delta_a= 0$ in  Algorithm 1 is the most desirable, and is used as the default value for our simulation study.

\subsubsection{Performance Comparison}
\begin{figure}[t]
\centering
\includegraphics[width=0.8\linewidth]{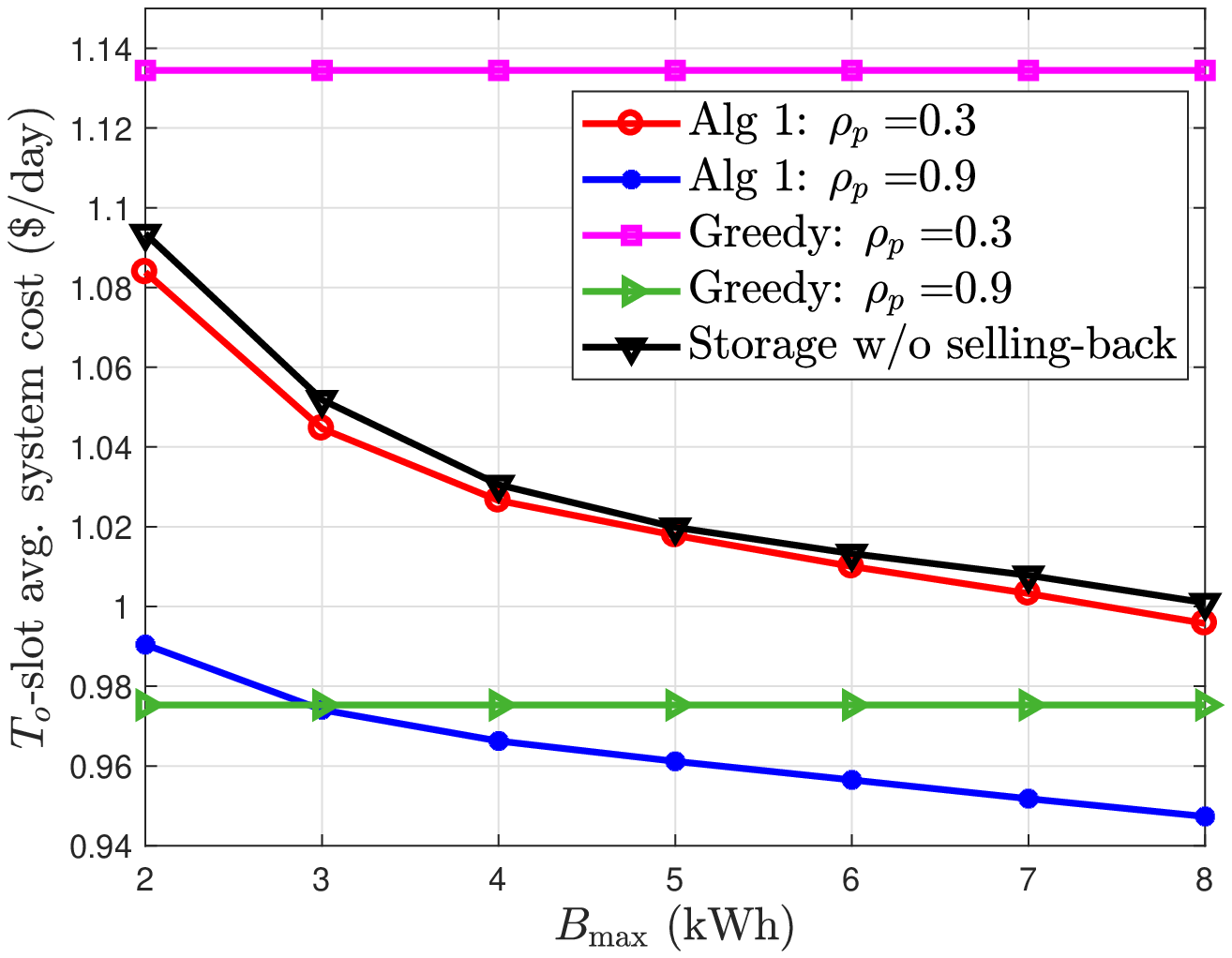}
\caption{\small Average system cost vs. battery capacity $B_{\max}$.}
\label{fig:Ymax}
\centering
\includegraphics[width=0.8\linewidth]{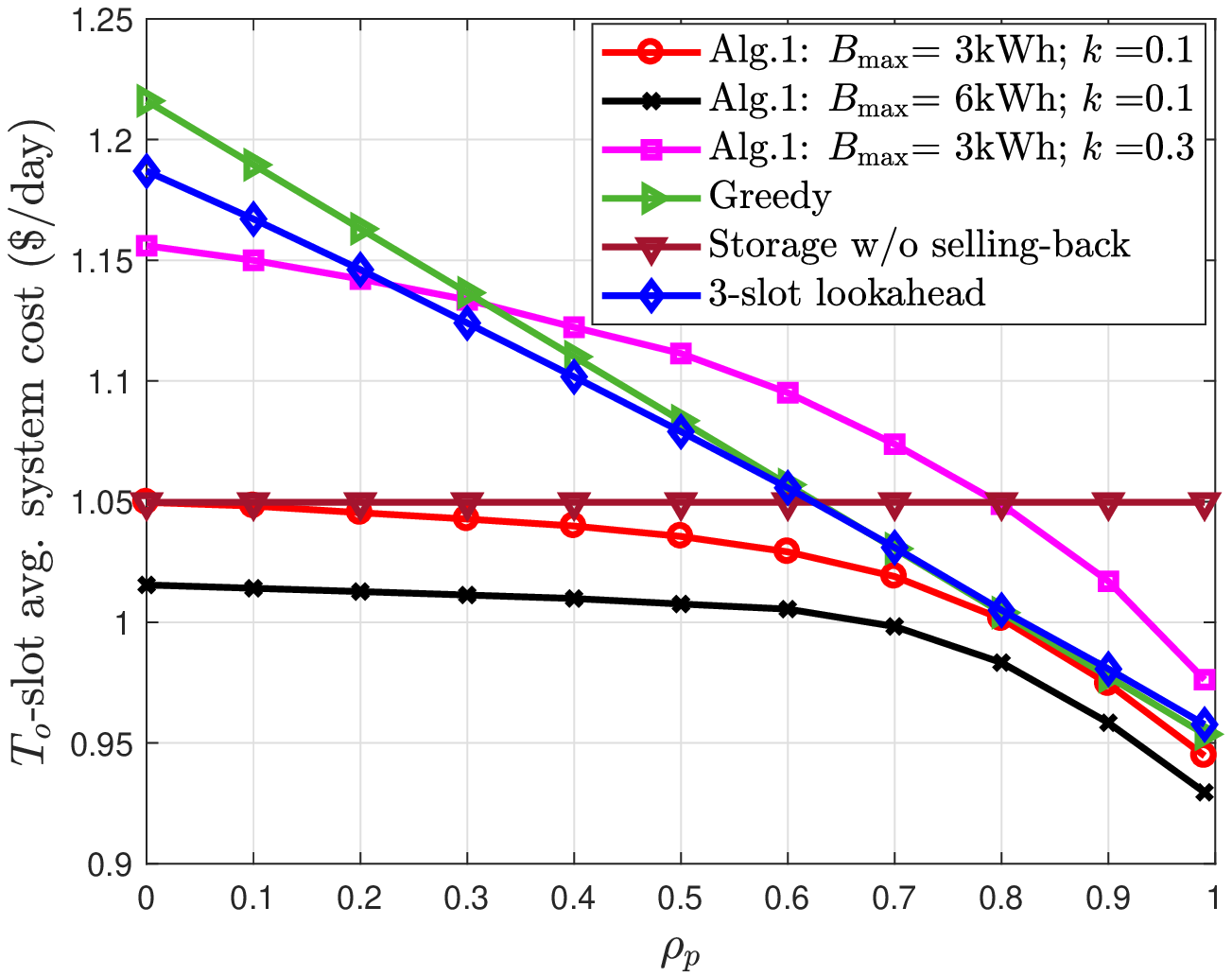}
\caption{\small Average system cost vs. selling-to-buying price ratio $\rho_p$ (Default: $B_{\max}=3$ kWh, $k=0.1$).}
\label{fig:percents_vs_cost}
\centering
\includegraphics[width=0.8\linewidth]{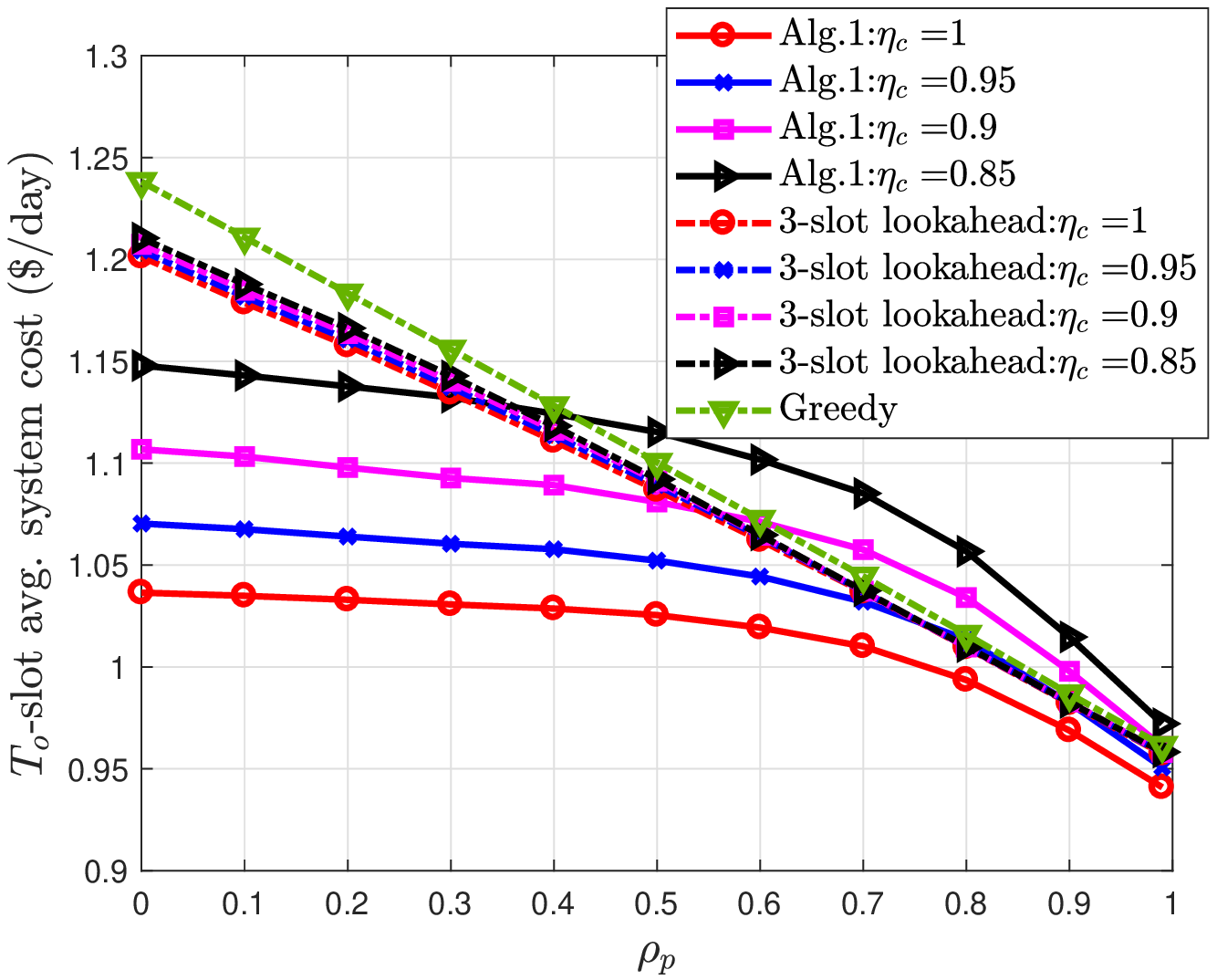}
\caption{\small Comparison of average system cost over $\rho_p$ at different battery charging (discharging) efficiency $\eta_c$ ($=\eta_d$).}
\label{fig:cost_vs_eta}
\end{figure}

We consider three other algorithms for comparison:
i) \emph{3-slot look-ahead}: The non-causal $T$-slot look-ahead optimal solution with $T=3$, where system inputs $\{\mubf_t\}$ for each 3-slot frame are known non-causally ahead of time. The resulting 3-slot minimum system cost is $u_m^\textrm{opt}$ for frame $m$.\footnote{The 3-slot look-ahead policy is obtained through exhaustive search. In general, the optimal solution can only be obtained from exhaustive search, which is difficult for larger $T$.}
ii) \emph{Greedy}: A greedy algorithm that minimizes the per-slot system cost, based on the current inputs $\mubf_t$. For one-slot cost minimization, to avoid the battery operation cost, the system directly buys energy from the grid to serve the load without storing energy into the battery. Thus, this greedy method is essentially a method without storage.
iii) \emph{Storage without selling-back}: With no energy selling-back capability in the model, the problem is reduced to the one considered in \cite{Li&Dong:JSAC15} and the algorithm proposed therein. Note that among the three methods, the first one is a non-causal approach and the rest   are real-time approaches.

 Fig.~\ref{fig:Ymax} compares  the average system cost of different algorithms at different  battery capacity $B_{\max}$, where
the  cost is converted into dollars per day. The system cost under Algorithm~1 reduces as $B_{\max}$ increases, because a larger battery capacity gives more flexibility on charging/discharging, allowing the system to buy more energy at a lower  price and sell more at a higher  price.
Also,  higher selling-to-buying price ratio $\rho_p=0.9$ results in a lower average system cost. For the  greedy algorithm, since it is essentially a method without the use of storage, the system cost does not vary with the battery capacity. Compared with the greedy algorithm, our proposed  Algorithm~1 offers more cost reduction as the battery capacity becomes larger.
Moreover, compared with the storage without selling-back, the extra cost saving by the availability to sell back energy at different selling price is clearly seen.

Fig.~\ref{fig:percents_vs_cost} provides the comparison of system cost  at various selling-to-buying price ratio $\rho_p$ values. For other algorithms, default $B_{\max}=3$ kWh and $k=0.1$ are used.
At the default setting, our proposed algorithm outperforms all other three methods for $\rho_p\in [0,1]$. By comparing Algorithm~\ref{alg1} with the storage without selling-back method, we see the cost saving by selling extra energy back to the grid   at different $\rho_p$. We also plot the performance of Algorithm~\ref{alg1} at a larger battery capacity and at a higher value of $k$ (higher battery usage cost) to see the effect of different battery specifications on the cost performance.

Finally, in Fig.~\ref{fig:cost_vs_eta}, we show the effect of different battery charging and discharging efficiencies on the performance, where we set $\eta_c=\eta_d$. Depending on the quality of battery, the efficiency may range from $0.85$ to $0.99$. As we see, compared to the 3-slot look-ahead solution, the increase of system cost  at lower $\eta_c(\eta_d)$ under Algorithm~\ref{alg1} is more noticeable. The performance of the greedy algorithm does not change with different $\eta_c$ since the algorithm doe not use the storage. When the charging efficiency is low ($0.85\sim0.9$) and  selling-to-buying price ratio is high, the greedy and 3-slot look-ahead methods can have a lower system cost, indicating that the benefit of storage diminishes in this scenario.

\section{Conclusion}\label{sec:conclusion}
In this work, we proposed a real-time bidirectional energy control algorithm for a residential ESM system to minimize the system cost within  a given time period. The ESM system has an integrated RG besides connected to the CG, and is capable of buy/sell energy from/to the CG. For the system cost, besides the energy buying cost and selling profit, we included the storage cost by accounting the battery operational cost and inefficiency due to charging/discharging activities. Our real-time algorithm provides a closed-form control solution for the ESM system which is very simple to implement. It does not relay on any statistical knowledge of system inputs and is applicable to arbitrary system input dynamics. We showed that our proposed algorithm provides a bounded performance guarantee to the non-causal $T$-slot look-ahead optimal control solution. Simulation demonstrated that our proposed algorithm outperforms other non-causal or real-time alternative methods. We further provided simulation study to understand how the availability of selling energy, selling and buying price setting, and battery inefficiency affect the storage behavior and system cost.

\appendices
\section{Equivalence of problems {\bf P2} and {\bf P3}}\label{appA}

The proof follows the general argument in \cite{Neely:ArXiv2010}.
The optimal solution of \textrm{\bf P2} satisfies all constraints of \textrm{\bf P3}, and therefore it is a feasible solution of {\bf P3}. Let $u^o_2$ and $u^o_3$ denote the minimum objective values of {\bf P2} and {\bf P3}, respectively. We have $u^o_3\le u^o_2$. By Jensen's inequality and convexity of $C(\cdot)$, we have
$\overline{C(\gamma)}\geq C(\overline{\gamma})=C(\overline{x_u})$, which means $u^o_3\ge u^o_2$. Hence,  $u^o_2=u^o_3$, and \textrm{\bf P2} and \textrm{\bf P3} are equivalent.

\section{Proof of Lemma~\ref{lemma drift upper bound}}\label{app drift upper bound}
\IEEEproof
By the definition of Lyapunov drift  $\Delta(\Thetabf_t)$,
\begin{align}\label{app:drift delta 1}
\Delta(\Thetabf_t)
&\triangleq L(\Thetabf_{t+1})-L(\Thetabf_t)=\frac{1}{2}\left(Z^2_{t+1}-Z^2_t+H^2_{t+1}-H^2_t\right)\nn\\
&=Z_t\left(\eta_c(Q_t+S_{c,t})-\frac{F_{d,t}+F_{s,t}}{\eta_d}-\frac{\Delta_a}{T_o}\right)\nn\\
&+H_t(\gamma_t-x_{u,t})+\frac{1}{2}(\gamma_t-x_{u,t})^2\nn\\
&+\frac{1}{2}\left(\eta_c(Q_t+S_{c,t})-\frac{F_{d,t}+F_{s,t}}{\eta_d}-\frac{\Delta_a}{T_o}\right)^2.
\end{align}

Let $g_t$ denote the sum of the last two terms in \eqref{app:drift delta 1}. From \eqref{eqn:delta_a}, we have $\frac{\Delta_a}{T_o}\le \max\{\eta_cR_{\max},D_{\max}/\eta_d\}=\Gamma$. For a given value of $\Delta_a$, by \eqref{eqn:rc_bds}, \eqref{eqn:dc_bds}, \eqref{eqn:x_2} and \eqref{eqn:gamma_bds}, $g_t$ is upper bounded by
\begin{align}\label{eqn:lemma_G}
g_t& \le \frac{\max\left\{\left(\eta_cR_{\max}-\frac{\Delta_a}{T_o}\right)^2,\left(\frac{D_{\max}}{\eta_d}+\frac{\Delta_a}{T_o}\right)^2\right\}}{2}+\frac{\Gamma^2}{2}
\triangleq G.
\end{align}

We now find the upper bound of $-H_tx_{u,t}$ in the second term of \eqref{app:drift delta 1}.

Note that $S_{w,t}$, $W_t$ and $H_t$ are known for the current time slot $t$. Also note that, $S_{w,t}- W_t \le 0$, because $S_{w,t}=\min\{W_t,S_t\}$.  The upper bound of $-H_tx_{u,t}$ is obtained as follows:
\newcounter{q1counter}
\begin{list}{{\it \arabic{q1counter}$\left.\right)$~}}
{\usecounter{q1counter}
\setlength\leftmargin{0em}
\setlength\labelwidth{0em}
\setlength\labelsep{0em}
\setlength\itemsep{0em}
}
\item {\it For $H_t\geq0$}: Let $l_t\triangleq S_{w,t}-W_t-F_{s,t}\le 0$. we have $x_{u,t}=|\eta_c(S_{c,t}+Q_t)-(F_{d,t}+F_{s,t})/\eta_d|\ge \eta_c|(S_{c,t}+Q_t)- (F_{d,t}-F_{s,t})|=\eta_c|E_t+S_{c,t}+S_{w,t}-W_t-F_{s,t}|$, where the last equality is by the supply-demand balance requirement in \eqref{eqn:Wt_constraint}. Thus,
\begin{align*}
-H_tx_{u,t}&\le-H_t\eta_c|E_t+S_{c,t}+S_{w,t}-W_t-F_{s,t}|\nn \\
&\leq-H_t\eta_c(|E_t+S_{c,t}|+S_{w,t}-W_t-F_{s,t}) \nn\\
&\leq-H_t\eta_c(E_t+S_{c,t}+l_t).
\end{align*}
\item {\it For $H_t<0$}: Let $l_t\triangleq W_t-S_{w,t}+F_{s,t}\ge 0$. We have $x_{u,t}\le|(S_{c,t}+Q_t)-(F_{d,t}+F_{s,t})|/\eta_d=|E_t+S_{c,t}+S_{w,t}-W_t-F_{s,t}|/\eta_d$. Thus,
\begin{align*}
-H_tx_{u,t}&\le-H_t|E_t-F_{s,t}+S_{c,t}+S_{w,t}-W_t|/\eta_d\\
&\leq -H_t(|E_t+S_{c,t}|+|S_{w,t}-W_t|+|F_{s,t}|)/\eta_d\\
&=-H_t/\eta_d(E_t+S_{c,t}+l_t).
\end{align*}
\end{list}
Finally, for the first term in \eqref{app:drift delta 1}, we have
$Z_t(\eta_c(Q_t+S_{c,t})-(F_{d,t}+F_{s,t})/{\eta_d}-\frac{\Delta_a}{T_o})\le Z_t (Q_t+S_{c,t}-F_{d,t}-F_{s,t}-\frac{\Delta_a}{T_o})=Z_t (E_t+S_{c,t}+S_{w,t}-W_t-F_{s,t}-\frac{\Delta_a}{T_o})$.
Combining the above results and \eqref{eqn:lemma_G}, we have the upper bound of $\Delta(\Thetabf_t)$ in \eqref{eqn:drift delta 1}.
\endIEEEproof

\section{Proof of Lemma~\ref{lemma33}}\label{app:lemma33}
\IEEEproof
The derivation follows the same steps as the one in Lemma 3 in \cite{Li&Dong:JSAC15}. We provide it here briefly. Since $C(\gamma_t)$ is a continuous, convex, non-decreasing function in $\gamma_t$ with $C'(\gamma_t)\ge 0$ and $C'(\gamma_t)$ increasing with $\gamma_t$. Denote the objective of $\text{\bf P4}_a$ by $J(\gamma_t)$. Since $\text{\bf P4}_a$ is convex, we examine the derivative of $J(\gamma_t)$ given by $J'(\gamma_t) = H_t+VC'(\gamma_t)$.

\newcounter{q2counter}
\begin{list}{{\it \arabic{q2counter}$\left.\right)$~}}
{\usecounter{q2counter}
\setlength\leftmargin{0em}
\setlength\labelwidth{0em}
\setlength\labelsep{0em}
\setlength\itemsep{0em}
}
\item {\it For $H_t\geq0$}: We have $J'(\gamma_t)>0$, thus $J(\gamma_t)$ monotonically increases, with its minimum obtained at $\gamma_t^*=0$.
\item {\it For $H_t<-VC'(\Gamma)$}: Since $VC'(\Gamma)\ge VC'(\gamma_t)$, we have $H_t+VC'(\gamma_t)<0$.  $J(\gamma_t)$ monotonically decreases, and its minimum is reached with $\gamma^*_t=\Gamma$.
\item  {\it For $-VC'(\Gamma)\le H_t\le 0$}: In this case, $\gamma^*_t$ is the root of $H_t+VC'(\gamma_t)=0$, given by $\gamma^*_t=C'^{-1}\left(-\frac{H_t}{V}\right)$.
\endIEEEproof
\end{list}

\section{Proof of Proposition~\ref{prop0}}\label{app:p4b}
\IEEEproof
Removing the constant terms in \eqref{drift_cost_metric} and from $l_t$ in \eqref{eqn:drift delta 1},  and after regrouping the rest of terms, we have the objective function of $\textbf{P4}_{\textbf{b}}$.

{\bf Determine $S^\w_{s,t}$ and $S^\w_{c,t}$}: To determine how to use the remaining renewable $S_t-S_{w,t}$, we need to minimize the term $S^\w_{c,t}(Z_t-g(H_t))+VC_{\textrm{rc}}-S^\w_{s,t}VP_{s,t}$ in the objective of $\text{\bf P4}_b$:
\begin{itemize}
\item[S1)] If $Z_t- g(H_t)> 0$: We should maximize $S^\w_{s,t}$ and minimize $S^\w_{c,t}$. Thus, the remaining amount should be sold back to the grid and not stored into the battery. We have $S^\w_{c,t}=0$, and $S^\w_{s,t}=\min\{S_t-S_{w,t},U_{\max}\}$ or $S^\w_{s,t}=\min\{S_t-S_{w,t}-F^\w_{s,t},U_{\max}\}$.
\item[S2)] If $Z_t-g(H_t)\le0$: The remaining renewable can be    stored into the battery and/or sold back to the grid. To minimize the cost, if $VP_{s,t}\ge {g(H_t)-Z_t}$, we set $S^\text{a}_{s,t}=\min\{S_t-S_{w,t},U_{\max}\}$,  $S^\text{a}_{c,t}=\min\{S_t-S_{w,t}-S^\w_{s,t},R_{\max}\}$, \ie first maximize the renewable sold back to the grid, then to the battery.  If $VP_{s,t}< {g(H_t)-Z_t}$, we set $S^\text{a}_{c,t}=\min\{S_t-S_{w,t},R_{\max}\}$,  $S^\text{a}_{s,t}=\min\{S_t-S_{w,t}-S^\w_{c,t},U_{\max}\}$, \ie first maximize the amount charged into the battery then consider the rest to be sold to the grid.
\end{itemize}

In the objective function  of $\text{\bf P4}_b$,  the optimal  $F_{s,t}$, $E_t$ and $S_{c,t}$ depend on the sign of $Z_t-|g(H_t)|+VP_{s,t}$, $Z_t-g(H_t)+VP_{b,t}$, and $Z_t-g(H_t)$, respectively. These three functions have the following relations: If $H_t\geq0$,
{
\begin{align}\label{eqn:relation h>0}
Z_t-g(H_t) \le Z_t-g(H_t)+VP_{s,t}<Z_t-g(H_t)+VP_{b,t}.
\end{align}
}
If $H_t<0$, the following two relations are possible
\begin{align}
&Z_t-g(H_t)\le Z_t-|g(H_t)|+VP_{s,t}<Z_t-g(H_t)+VP_{b,t},\label{eqn:relation h<0 1}\\
&Z_t-|g(H_t)|+VP_{s,t}\le Z_t-g(H_t)<Z_t-g(H_t)+VP_{b,t}.\label{eqn:relation h<0 2}
\end{align}
Based on the relations in \eqref{eqn:relation h>0}--\eqref{eqn:relation h<0 2}, we have the following five cases and derive the optimal solution in each case.
\subsubsection{For $Z_t-g(H_t)+VP_{b,t}\leq0$} From \eqref{eqn:relation h>0} and \eqref{eqn:relation h<0 2}, to minimize the objective function of   $\text{\bf P4}_b$, $F_{s,t}=0$, and we want to maximize $E^\w_t$.
This means the battery is in the charging state. We have $1_{R,t}=1$,  $1_{D,t}=0$, $F^\w_{d,t}=F^\w_{s,t}=0$, and use maximum charging rate $S^\w_{c,t}+Q^\w_t=R_{\max}$ if possible.
Since $Z_t-g(H_t)\leq Z_t-g(H_t)+VP_{b,t}\le 0$, between $Q^\w_t$ and $S^\w_{c,t}$,  determining $S^\w_{c,t}$ first will further reduce the objective value of $\text{\bf P4}_b$. Since $Z_t-g(H_t)\le 0$,  $(S^\w_{c,t},S^\w_{s,t})=(S^\text{a}_{c,t},S^\text{a}_{s,t})$ as in S2) given earlier.  By supply-demand balancing equation \eqref{eqn:Wt_constraint}, we obtain $Q^\w_t$ and $E^\w_t$ in \eqref{case a}.
Alternatively, we can keep the battery idle and only buy energy $E_t^\textrm{id}$ from the grid, where $S_{c,t}^\textrm{id}+Q_t^\textrm{id}=0$. In this case,  the battery cost can be avoided: $1_{R,t}=1_{D,t}=0$, but $E_t^\textrm{id}$ will be smaller. The optimal $\abf^*_t$ is the one that achieves the minimum objective value.

\subsubsection{For $\max\{Z_t-g(H_t),Z_t-|g(H_t)|+VP_{s,t}\}<0\leq Z_t-g(H_t)+VP_{b,t}$} In this case, to minimize the objective of  $\text{\bf P4}_b$, we want to set $E^\w_t$ as small as possible and $F^\w_{s,t}=0$. It is possible that the battery is in either charging or discharging state. If we charge the battery, it should only be charged from renewable $S^\w_{c,t}$, while $Q^\w_t=0$. Since $Z_t-g(H_t)< 0$, $(S^\w_{c,t},S^\w_{s,t})=(S^\text{a}_{c,t},S^\text{a}_{s,t})$  as in S2) given earlier. Note that, if in the charging state, it means $S^\w_{c,t}>0$, which happens when  $W_t=S_{w,t}<S_t$, and we have   $F^\w_{d,t}=0$. Thus, constraint \eqref{eqn:rc_dc_constr} is satisfied.  On the other hand, if $W_t>S_{w,t}$, it means $S_t=S_{w,t}$ and $S^\w_{s,t}=S^\w_{c,t}=0$. We could either use the battery and/or buy energy  $E^\w_t$ (\ie idle or discharging state) to serve the load. If the battery is in the discharging state, the amount  $F^\w_{d,t}$ should be set as large as possible to minimize  $E^\w_t$.  Based on the above, we have the control solution $\abf^\w_t$ as shown in \eqref{case c}.
Alternatively, we can keep the battery idle to avoid battery cost, and only buy energy $E_t^\textrm{id}$ from the grid. Thus, the optimal $\abf^*_t$ is chosen by whichever achieves the minimum objective value.

\subsubsection{For $Z_t-g(H_t) \le 0\le Z_t-|g(H_t)|+VP_{s,t}$} This is the case when \eqref{eqn:relation h>0} and \eqref{eqn:relation h<0 1} hold.
To minimize the objective of $\text{\bf P4}_b$, one possible solution is to minimize $E^\w_t$ and maximize $F^\w_{s,t}$. Due to constraint \eqref{eqn:rc_dc_constr}, $S^\w_{c,t}\cdot F^\w_{s,t}=0$ must be satisfied. Thus, we have two conditions: i) If $F^\w_{s,t}\geq0$ and $S^\w_{c,t}=0$, it means the remaining amount $S_t-S_{w,t}$, if any, will be only sold back to the grid. Since $Z_t\le g(H_t)$, it is easy to see that $0<Z_t-|g(H_t)|+VP_{s,t}\le VP_{s,t}$. Thus, we first maximize  $S^\w_{s,t}$ and then maximize $F^\w_{s,t}$. Since $Z_t-g(H_t)\le 0$, $(S^\w_{c,t},S^\w_{s,t})=(S^\text{a}_{c,t},S^\text{a}_{s,t})$  as in S2). The control solution $\abf_t^\w$ is shown as in \eqref{case d 2}.
ii) If $S^\w_{c,t}\geq0$ and $F^\w_{s,t}=0$, the battery will be charged from $S^\w_{c,t}$ only and no energy from the battery will be sold. We have $(S^\w_{c,t},S^\w_{s,t})=(S^\text{a}_{c,t},S^\text{a}_{s,t})$  as in S2). The control solution $\abf^\w_t$ is shown as in \eqref{case d 3}. After comparing i) and ii), $\abf^\w_t$ is the one whichever achieves the less objective value.
Alternatively, we can keep the battery idle. Thus, the optimal $\abf^*_t$ is chosen by whichever achieves the minimum objective value between $\abf^\w_t$ and $\abf_t^\textrm{id}$.

\subsubsection{For $Z_t-|g(H_t)|+VP_{s,t}<0 \le Z_t-g(H_t)$} Note that this case  happens when $H_t<0$ and \eqref{eqn:relation h<0 2} holds.
We want to set  $E^\w_t$ as small as possible and $F^\w_{s,t}=0$. Since $Z_t\ge g(H_t)$, from earlier we have  $S^\w_{c,t}=0$. Thus, the battery  can be in the discharging state, and it is straightforward to obtain $\abf^\w_t$ in \eqref{case f}. After comparing to the alternative idle state, the optimal $\abf^*_t$ is chosen by whichever achieves the minimum objective value.
\subsubsection{For $\min\{Z_t-g(H_t),Z_t-|g(H_t)|+VP_{s,t}\}>0$} Since $Z_t> g(H_t)$,  $S^\w_{c,t}=0$. By \eqref{eqn:relation h>0}-\eqref{eqn:relation h<0 2}, to minimize the objective of $\text{\bf P4}_b$,  we want to minimize $E^\w_t$ and maximize $F^\w_{s,t}$. This means no charging:  $Q^\w_t=0$.
Thus, only the discharging  or idle state could be considered.
 For the discharging state, since $Z_t-|g(H_t)|+VP_{s,t}<Z_t-g(H_t)+VP_{b,t}$, we should first maximize the discharging amount as $F^\w_{d,t}=\min\{W_t-S_{w,t},D_{\max}\}$ to minimize $E^\w_t$, then maximize  $F^\w_{s,t}$. For energy selling, between $F^\w_{s,t}$ and $S^\w_{s,t}$, to minimize the cost, if $Z_t-|g(H_t)|>0$, we should first maximize $F^\w_{s,t}$ from the battery and then sell from the renewable, as $F^\w_{s,t}$ and $S^\w_{s,t}$ in \eqref{case b 1}. Otherwise, we first sell as much as possible from the renewable, and then determine    $F^\w_{s,t}$ as given in \eqref{case b 2}. By supply-demand  equation \eqref{eqn:Wt_constraint},  $E^\w_t$ can be obtained as in \eqref{case b 1} and \eqref{case b 2}.
Alternatively, we can keep the battery idle and only buy energy $E_t^\textrm{id}$ from the grid. The optimal $\abf^*_t$ is the one that achieves the minimum objective value.

\section{Proof of Proposition~\ref{prop1}}\label{appC}
\IEEEproof
Before going into the details, we first provide an outline of our proof. Using the solutions $\gamma^*_t$ and $\mathbf{a}^*_t$ of  $\text{\bf P4}_a$ and  $\text{\bf P4}_b$, respectively, we can show that both $Z_t$ and $H_t$ are upper and lower bounded. Then, by applying these bounds to \eqref{eqn:dynamic shift Z} and using the battery capacity constraint $\eqref{eqn:Yt bds}$, we obtain $A_o$ as the minimum value that can be achieved with a given value of $\Delta_a$. With $A_o$ obtained, we derive the upper bound of $V$, \ie $V_{\max}$, to ensure that $\eqref{eqn:Yt bds}$ is satisfied.

To prove Proposition~\ref{prop1}, we first introduce Lemma~\ref{app lemma of H} and Lemma~\ref{app lemma bound of Z} below.
\begin{lemma}\label{app lemma of H}
For $\gamma^*_t$ in \eqref{eqn:optimal gamma}, $H_t$ is bounded by
\begin{align}\label{eqn:app H bound}
H_{\min} \le H_t \le H_{\max}
\end{align}
where $H_{\min}\triangleq-VC'(\Gamma)-\Gamma<0$, and $H_{\max}\triangleq\Gamma$.
\end{lemma}
\IEEEproof
\setcounter{subsubsection}{0}
\subsubsection{Upper bound}
From \eqref{eqn:x_2}, $x_{u,t}\ge 0$. If $H_t\ge 0$, from \eqref{eqn:optimal gamma}, we have $\gamma^*_t=0$. Thus, based on the dynamics of $H_t$ in \eqref{eqn:queue H}, $H_{t+1}\le H_t$, \ie non-increasing. When $H_t < 0$, from \eqref{eqn:optimal gamma}, the maximum increment of $H_{t+1}$ in \eqref{eqn:queue H} is when $\gamma^*_t=\Gamma$ and $x_{u,t}=0$, and thus $H_{t+1} \le \Gamma=H_{\max}$.
\subsubsection{Lower bound}
From \eqref{eqn:optimal gamma}, if $H_t<-VC'(\Gamma)$, we have $\gamma_t^*=\Gamma$, and $H_{t+1}$ is non-decreasing in \eqref{eqn:queue H}. If $H_t\ge -VC'(\Gamma)$, the maximum decrement of $H_{t+1}$ from $H_t$ in \eqref{eqn:queue H} is when $\gamma_t^*=0$ and $x_{u,t}=\Gamma$, and $H_{t+1} \ge -VC'(\Gamma)-\Gamma=H_{\min}$. Since $C(\cdot)$ is non-decreasing, $C'(\cdot)\ge0$, we have $H_{\min}<0$.
\endIEEEproof

\begin{lemma}\label{app lemma bound of Z}
Under the proposed solution in Proposition~\ref{prop0}, we have\\
1) If $Z_t<-VP_{b}^{\max}+H_{\min}/\eta_d$, then $F^*_{d,t}+F^*_{s,t}=0$; \\
2) If $Z_t>\max\{\eta_c H_{\max}, |H_{\min}|/\eta_d-VP_{s}^{\min}\}$, then $S^*_{r,t}+Q^*_t=0$.
\end{lemma}

\IEEEproof
Note that $H_{\min}< 0$ and $H_{\max}>0$.
 1) This case corresponds to~Case 1) in Proposition~\ref{prop0}.
We have $Z_t<-VP_{b}^{\max}+H_{\min}/\eta_d\le -VP_{b}^{\max}+g(H_t)$, thus $F^*_{d,t}+F^*_{s,t}=0$ is the optimal control action.
2) This case corresponds to Case~5) in Proposition~\ref{prop0}.
From Lemma~\ref{app lemma of H}, we know $|H_{\min}|>|H_{\max}|$.
Thus, it is easy to see that if $Z_t>\max\{\eta_cH_{\max}, |H_{\min}|/\eta_d-VP_{s}^{\min}\}$, then $S_{c,t}^*=Q^*_t=0$ are the optimal control action.
\endIEEEproof
Now, we are ready to prove Proposition~\ref{prop1}.  When first show that under $A_o$ and $V$ in \eqref{eqn:A_o} and \eqref{eqn:V_max}, $B_t$ is always upper bounded by $B_{\max}$; Then we prove that $B_t$ is lower bounded by $B_{\min}$.
\subsection{Upper Bound  $B_t\leq B_{\max}$}
Based on Lemma \ref{app lemma bound of Z}.1), we have $F^*_{d,t}+F^*_{s,t}=0$ if $Z_t<-VP_{b}^{\max}+ H_{\min}/\eta_d$, no discharging from the battery.
When $Z_t\geq -VP_{b}^{\max}+H_{\min}/\eta_d$, from \eqref{eqn:dc_bds},  the maximum decreasing amount  from $Z_t$ to $Z_{t+1}$ in \eqref{eqn:queue Z} in the next time slot is $D_{\max}/\eta_d$, and we have, for $t\in [0,T_o-1]$,
\begin{align}\label{eqn:Z_{t+1}_lower bound}
Z_{t+1}\geq -VP_{b}^{\max}-\frac{\Delta_a}{T_o}-\!\frac{D_{\max}\!-\!H_{\min}}{\eta_d}.
\end{align}
In \eqref{eqn:dynamic shift Z}, we have $B_t=Z_t+A_o+\frac{\Delta_a}{T_o}t$. To satisfy the lower bound of $B_t$ in \eqref{eqn:Yt bds}, we must ensure $Z_t+A_o+\frac{\Delta_a}{T_o}t\geq B_{\min}$. From \eqref{eqn:Z_{t+1}_lower bound}, it is sufficient  to let
$-VP_{b}^{\max}-\frac{\Delta_a}{T_o}-(D_{\max}-H_{\min})/\eta_d+A_o+\frac{\Delta_a}{T_o}t\geq B_{\min}$,
which means
\begin{align}\label{eqn:app A_o}
A_o\geq  B_{\min}+VP_{b}^{\max}\!+\!\frac{D_{\max}\!-\!H_{\min}}{\eta_d}+\frac{\Delta_a}{T_o}(1-t),
\end{align}
for all  $t\in [0, T_o]$.
We next determine the minimum possible value of $A_o$ based on the sign of $\Delta_a$.
\setcounter{subsubsection}{0}
\subsubsection{If $\Delta_a\geq0$} The minimum value of $A_o$ in \eqref{eqn:app A_o} is
\begin{align}\label{eqn:app A_o 1}
\hspace*{-.5em}A_{o,\min}&=B_{\min}+VP_{b}^{\max}+\frac{D_{\max}\!-\!H_{\min}}{\eta_d}+\frac{\Delta_a}{T_o}.
\end{align}
As a result, we have $A_t=A_{o,\min}+\frac{\Delta_a}{T_o}t$.

Based on Lemma \ref{app lemma bound of Z}.2), we have $S^*_{r,t}+Q^*_t=0$ if  $Z_t-\frac{\Delta_a}{T_o}>\max\{\eta_cH_{\max},|H_{\min}|/\eta_d-VP_{s}^{\min}\}-\frac{\Delta_a}{T_o}$, \ie no charging for the battery.
When $Z_t-\frac{\Delta_a}{T_o}\leq \max\{\eta_cH_{\max},|H_{\min}|/\eta_d-VP_{s}^{\min}\}-\frac{\Delta_a}{T_o}$, based on the maximum increment of $Z_t$ to $Z_{t+1}$ in \eqref{eqn:queue Z}, we have, for $t\in[0,T_o]$,
\begin{align}\label{eqn:Z_{t+1} upper bound}
\hspace*{-.5em}Z_t\leq \max\{\eta_cH_{\max},\frac{|H_{\min}|}{\eta_d}\!-\!VP_{s}^{\min}\}\!-\frac{\Delta_a}{T_o}\!+\eta_c R_{\max}.
\end{align}
Substituting $A_t$ with  $A_{o,\min}$ in \eqref{eqn:app A_o 1} into \eqref{eqn:dynamic shift Z}, and from \eqref{eqn:Z_{t+1} upper bound}, we have
\begin{align}\label{eqn:app2_Y_1}
B_t\leq& B_{\min}+\max\{\eta_cH_{\max},\frac{|H_{\min}|}{\eta_d}\!-VP_{s}^{\min}\}+\eta_cR_{\max}\nn\\
&+VP_{b}^{\max}+\frac{D_{\max}\!-\!H_{\min}}{\eta_d}+\frac{\Delta_a}{T_o}t.
\end{align}
For the control solution to be feasible, we need $B_t\le B_{\max}$. This is satisfied if RHS of \eqref{eqn:app2_Y_1} $\le B_{\max}$, for all $t\in [0,T_o]$. Using $H_{\min}$ and $H_{\max}$ expressions in  Lemma~\ref{app lemma of H}, this means
\begin{align*}
&VP_b^{\max} \le \tilde{B}_t+\frac{H_{\min}}{\eta_d}-\max\{\eta_cH_{\max},\frac{|H_{\min}|}{\eta_d}-VP_{s}^{\min}\} \nn \\
&\le \tilde{B}_t-\frac{VC'(\Gamma)}{\eta_d}-\Gamma(\frac{1}{\eta_d}+\eta_c)\nn \\
&-\max\{0,\ V(\frac{C'(\Gamma)}{\eta_d}-P_{s}^{\min})+\Gamma(\frac{1}{\eta_d}-\eta_c)\} \nn \\
&=\!  \tilde{B}_t\!-\!\frac{VC'(\Gamma)\!-\!2\Gamma }{\eta_d}\!-\!\max\{\Gamma(\eta_c\!-\!\frac{1}{\eta_d}),  V(\frac{C'(\Gamma)}{\eta_d}\!-\!P_{s}^{\min})\}
\end{align*}
where $\tilde{B}_t\triangleq \! B_{\max}\!-B_{\min}\!-\eta_cR_{\max}\!-D_{\max}/\eta_d-\frac{\Delta_a}{T_o}t$.
To satisfy the above inequality, it is suffice that the following holds
\begin{align*}
VP_b^{\max} &\le\tilde{B}_t-\frac{VC'(\Gamma)-2\Gamma }{\eta_d}-V\max\{0,  (\frac{C'(\Gamma)}{\eta_d}-P_{s}^{\min})\},
\end{align*}
which is satisfied if $V\in (0,V_{\max}]$ with
\begin{align}\label{eqn:Vmax for delta_a>0}
V_{\max}=\frac{\tilde{B}_0-2\Gamma/\eta_d-\Delta_a}{P_{b}^{\max}+C'(\Gamma)/\eta_d+[C'(\Gamma)/\eta_d-P_{s}^{\min}]^+}.
\end{align}
\subsubsection{If $\Delta_a<0$} The minimum value of $A_o$ in \eqref{eqn:app A_o} is \begin{align}\label{eqn:app A_o 2}
\hspace*{-.5em}A_{o,\min}=\!B_{\min}\!+\!VP_{b}^{\max}+\frac{D_{\max}-H_{\min}}{\eta_d}+\frac{\Delta_a}{T_o}-\Delta_a.
\end{align}
Substituting $A_{o,\min}$ in $A_t$, and  from \eqref{eqn:dynamic shift Z} and  \eqref{eqn:Z_{t+1} upper bound}, we have
\begin{align}\label{eqn:app2_Y_2}
B_t\leq& B_{\min}+\max\{\eta_cH_{\max},\frac{|H_{\min}|}{\eta_d}\!-VP_{s}^{\min}\}+\eta_cR_{\max}\nn\\
&+VP_{b}^{\max}+\frac{D_{\max}\!-\!H_{\min}}{\eta_d}-\Delta_a+\frac{\Delta_a}{T_o}t.
\end{align}
Again, to satisfy $B_t\le B_{\max}$, it is suffice that RHS of \eqref{eqn:app2_Y_2} $\le B_{\max}$. Using the similar analysis in the case of $\Delta_a\ge0$ , the bound  is satisfied if $V\in (0,V_{\max}]$ with
\begin{align}\label{eqn:Vmax for delta_a<0}
V_{\max}=\frac{\tilde{B}_0-2\Gamma/\eta_d-|\Delta_a|}{P_{b}^{\max}+C'(\Gamma)/\eta_d+[C'(\Gamma)/\eta_d-P_{s}^{\min}]^+}.
\end{align}
Combining \eqref{eqn:Vmax for delta_a>0} and \eqref{eqn:Vmax for delta_a<0} leads to  \eqref{eqn:V_max}, and from  \eqref{eqn:app A_o 1} or \eqref{eqn:app A_o 2}, we have $A_o$ in \eqref{eqn:A_o}.

\subsection{Lower Bound  $B_t\geq B_{\min}.$}
We now show that using $A_{o,\min}$ in \eqref{eqn:app A_o 1} or \eqref{eqn:app A_o 2} for $\Delta_a\ge 0$ or $\Delta_a<0$, respectively, and $V\in (0,V_{\max}]$ with $V_{\max}$ in \eqref{eqn:Vmax for delta_a>0} or \eqref{eqn:Vmax for delta_a<0}, respectively, we have $B_t\ge B_{\min}$ for all $t$.
\setcounter{subsubsection}{0}
\subsubsection{If $\Delta_a\geq0$} Substitute $A_{o,\min}$ in \eqref{eqn:app A_o 1} in $A_t$, and $Z_t$ in \eqref{eqn:dynamic shift Z} into \eqref{eqn:Z_{t+1}_lower bound}, we have
 $B_{\min}+\frac{\Delta_a}{T_o}t\leq B_t$. Since $\frac{\Delta_a}{T_o}t>0$, for $t\in [0,T_o-1]$, $B_t\geq B_{\min}$ is satisfied for $\Delta_a\geq0$.

\subsubsection{If $\Delta_a<0$}
Substitute $A_{o,\min}$ in \eqref{eqn:app A_o 2} in $A_t$, and $Z_t$ in \eqref{eqn:dynamic shift Z} into \eqref{eqn:Z_{t+1}_lower bound}, we have
 $B_{\min}+\Delta_a(\frac{t}{T_o}-1)\leq B_t$.
Since $\Delta_a(\frac{t}{T_o}-1)>0$, for $t\in [0,T_o-1]$,  $B_t\geq B_{\min}$ is satisfied for $\Delta_a<0$.
\endIEEEproof

\section{Proof of Theorem~\ref{thm1}}\label{appD}
\IEEEproof
A $T$-slot sample path Lyapunov drift is defined by $\Delta_T(\Theta_t)\triangleq L(\Thetabf_{t+T})-L(\Thetabf_{t})$. We upper bound it as follows
{
\allowdisplaybreaks
\begin{align}\label{eqn:app3_Delta_T}
&\Delta_T(\Theta_t)
=\frac{1}{2}\left( Z^2_{t+T}-Z^2_{t}+H^2_{t+T}-H^2_{t}\right)\nn\\
&=Z_t\sum_{\tau=t}^{t+T-1}\left( \eta_c(Q_\tau+S_{r,\tau})-(F_{d,\tau}+F_{s,\tau})/\eta_d-\frac{\Delta_a}{T_o}\right)\nn\\
&\ +H_t\sum_{\tau=t}^{t+T-1}\left(\gamma_\tau-x_{u,\tau}\right)+\frac{1}{2}\left[\sum_{\tau=t}^{t+T-1}\left(\gamma_\tau-x_{u,\tau}\right)\right]^2\nn\\
&\quad+\frac{1}{2}\left[\sum_{\tau=t}^{t+T-1}\left(\eta_c(Q_\tau+S_{r,\tau})-(F_{d,\tau}+F_{s,\tau})/\eta_d-\frac{\Delta_a}{T_o}\right)\right]^2\nn\\
&\leq Z_t\sum_{\tau=t}^{t+T-1}\left(\eta_c(Q_\tau+S_{r,\tau})-(F_{d,\tau}+F_{s,\tau})/\eta_d-\frac{\Delta_a}{T_o}\right)\nn\\
&\quad+H_t\sum_{\tau=t}^{t+T-1}(\gamma_\tau-x_{u,\tau})+G T^2
\end{align}
}where $G$ is defined in Lemma~\ref{lemma drift upper bound}.

Let $T_o=MT$. We consider a per-frame optimization problem below,
with the objective of minimizing the time-averaged system cost within the $m$th frame of length $T$ time slots.
\begin{align}
{\bf \textrm{\bf P}_f:} \; &\min_{\{\abf_t,\gamma_t\}} \;
\frac{1}{T}\sum_{t=mT}^{(m+1)T-1}[E_tP_{b,t}-(F_{s,t}+S_{s,t})P_{s,t}\nn\\
&\hspace{9em}+x_{e,t}+C(\gamma_t)]\nn\\
\rm{s.t} \;\;
&\eqref{eqn:Wt_constraint},\eqref{eqn:Solar S2 S3 bounds},\eqref{eqn:rc_dc_constr},
\eqref{eqn:rc_bds_strict},\eqref{eqn:dc_bds_strict},\eqref{avg_r=avg_x},\ \textnormal{and}\ \eqref{eqn:gamma_bds}.\nn
\end{align}

We show that ${\bf \textrm{\bf P}_f}$ is equivalent to {\bf P1} in which $T_o$ is replaced by $T$.
Let $u_m^f$ denote the minimum objective value of ${\bf \textrm{\bf P}_f}$. The optimal solution of {\bf P1} satisfies all constraints of ${\bf \textrm{\bf P}_f}$ and therefore is feasible to ${\bf \textrm{\bf P}_f}$. Thus, we have $u_m^f\leq u_m^\textrm{opt}$. By Jensen's inequality and convexity of $C(\cdot)$, we have $\overline{C(\gamma)}\geq C(\overline{\gamma})=C(\overline{x_u})$.
Note that introducing the auxiliary variable $\gamma_t$ with constraints \eqref{eqn:gamma_bds} and \eqref{avg_r=avg_x} does not modify the problem.
This means $u_m^f\ge u_m^\textrm{opt}$.
Hence, we have $u_m^f= u_m^\textrm{opt}$ and ${\bf \textrm{\bf P}_f}$ and {\bf P1} are equivalent.

From \eqref{eqn:app3_Delta_T} and the objective of ${\bf \textrm{\bf P}_f}$, we have the $T$-slot drift-plus-cost metric for the $m$th frame upper bounded by
\begin{align}\label{eqn:T slot drift plus penalty}
&\Delta_T(\Theta_t)\nn\\
& +V\!\!\!\sum_{t=mT}^{(m+1)T-1}[E_tP_{b,t}-(F_{s,t}+S_{s,t})P_{s,t}+x_{e,t}+C(\gamma_t)]\nn\\
&\leq Z_{t\!\!\!}\sum_{t=mT}^{(m+1)T-1}\!\!\!\left(\eta_c(Q_\tau+S_{r,\tau})-(F_{d,\tau}+F_{s,\tau})/\eta_d-\frac{\Delta_a}{T_o}\right)\nn\\
&\quad+H_{t}\!\!\!\sum_{t=mT}^{(m+1)T-1}\!\!\!\left(\gamma_t-x_{u,t}\right)+G T^2\nn\\
&+V\!\!\!\!\!\sum_{t=mT}^{(m+1)T-1}\!\!\!\!\![E_tP_{b,t}-(F_{s,t}+S_{s,t})P_{s,t}+x_{e,t}+C(\gamma_t)].
\end{align}
Let $\{\tilde{\abf}_t,\tilde{\gamma}_t\}$ denote a pair of feasible solution of ${\bf \textrm{\bf P}_f}$, satisfying the following relations
\begin{align}
&\hspace*{-.7em}\sum_{t=mT}^{(m+1)T-1}\!\!\!\!\!\! \eta_c\left(\tilde{Q}_t+\tilde{S}_{r,t}\right)=\!\!\!\sum_{t=mT}^{(m+1)T-1}\!\!\left(\frac{\tilde{F}_{d,t}+\tilde{F}_{s,t}}{\eta_d}+\frac{\Delta_a}{T_o}\right)\label{eqn:avg_rc_dc_bds_per_frame_1}\\
&\hspace*{-.7em}\sum_{t=mT}^{(m+1)T-1}\!\!\!\tilde{\gamma}_t=\sum_{t=mT}^{(m+1)T-1}\!\!\!\tilde{x}_{u,t}\label{eqn:avg_rc_dc_bds_per_frame_2}
\end{align}
with the corresponding objective value denoted as $\tilde{u}^f_m$.

Note that comparing with {\bf P1}, we impose per-frame constraints \eqref{eqn:avg_rc_dc_bds_per_frame_1} and \eqref{eqn:avg_rc_dc_bds_per_frame_2} as oppose to \eqref{eqn:delta_a} and \eqref{avg_r=avg_x} for the $T_o$-slot period.
Let $\delta\geq 0$ denote the gap of $\tilde{u}^f_m$ to the optimal objective value $u_m^\textrm{opt}$, \ie $\tilde{u}^f_m=u_m^\textrm{opt}+\delta$.

Among all feasible control solutions satisfying \eqref{eqn:avg_rc_dc_bds_per_frame_1} and \eqref{eqn:avg_rc_dc_bds_per_frame_2}, there exists a solution which leads to $\delta\rightarrow0$. The upper bound in \eqref{eqn:T slot drift plus penalty} can be rewritten as
\begin{align}\label{eqn:T slot drift plus penalty, per frame}
&\Delta_T(\Theta_t) +V\!\!\!\!\!\sum_{t=mT}^{(m+1)T-1}\!\!\!\!\![E_tP_{b,t}-(F_{s,t}+S_{s,t})P_{s,t}+x_{e,t}+C(\gamma_t)]\nn\\
&\leq G T^2+VT\lim_{\delta\rightarrow0}\left(u_m^\textrm{opt}+\delta\right)= G T^2+VTu_m^\textrm{opt}.
\end{align}
Summing both sides of \eqref{eqn:T slot drift plus penalty, per frame} over $m$ for $m=0,\ldots, M-1$, and dividing them by $VMT$, we have
{\small
\begin{align}\label{eqn:T slot drift plus penalty, 2}
&\frac{L(\Thetabf_{T_o})-L(\Thetabf_0)}{VMT}\nn\\
&\ +\frac{1}{MT}\sum_{m=0}^{M-1}\sum_{t=mT}^{(m+1)T-1}\!\!\!\!\![E_tP_{b,t}-(F_{s,t}+S_{s,t})P_{s,t}+x_{e,t}+C(\gamma_t)]\nn\\
&\leq
\frac{1}{M}\sum_{m=0}^{M-1}u_m^\textrm{opt}+\frac{G T}{V}.
\end{align}}

Since $\overline{C(\gamma)}\geq C(\overline{\gamma})$ for the convex function $C(\cdot)$ where $\overline{\gamma}\triangleq \frac{1}{T_o}\sum_{t=0}^{T_o-1}\gamma_t$, from \eqref{eqn:T slot drift plus penalty, 2}, we have
\begin{align}\label{eqn:T slot drift plus penalty, 3}
&\hspace*{-0.5em}\frac{1}{T_o}\sum_{t=0}^{T_o-1}[E_tP_{b,t}-(F_{s,t}+S_{s,t})P_{s,t}]+\overline{x_e}+C(\overline{\gamma})\nn\\
&\hspace*{-0.5em}\leq
\frac{1}{T_o}\sum_{t=0}^{T_o-1}[E_tP_{b,t}-(F_{s,t}+S_{s,t})P_{s,t}+x_{e,t}+C(\gamma_t)].
\end{align}
For a continuously differentiable convex function $f(\cdot)$, the following inequality holds \cite{book:Boyd&Vandenberghe}
\begin{align}\label{eqn:app3 convex}
f(x)\geq f(y)+f'(y)(x-y).
\end{align}
Applying \eqref{eqn:app3 convex} to $C(\overline{x_u})$ and $C(\overline{\gamma})$, we have
\begin{align}\label{eqn:app3 convex2}
C(\overline{x_u})
&\leq C(\overline{\gamma})+C'(\overline{x_u})(\overline{x_u}-\overline{\gamma})
\leq C(\overline{\gamma})+C'(\Gamma)(\overline{x_u}-\overline{\gamma})\nn\\
&=C(\overline{\gamma})-C'(\Gamma)\frac{H_{T_o}-H_0}{T_o}
\end{align}
where the last term in \eqref{eqn:app3 convex2} is obtained by summing both sides of \eqref{eqn:queue H} over $T_o$.

Applying the inequality \eqref{eqn:app3 convex2} to $C(\overline{\gamma})$ at the LHS of \eqref{eqn:T slot drift plus penalty, 3}, and further applying the inequality \eqref{eqn:T slot drift plus penalty, 3} to the LHS of \eqref{eqn:T slot drift plus penalty, 2}, we have the  bound of the objective value $u^*(V)$ of {\bf P1} achieved by Algorithm~\ref{alg1}
as in \eqref{thm1:bd}.

For the bound in \eqref{thm1:bd}, note that $H_t$ is bounded as in \eqref{eqn:app H bound}, and $Z_t$ is bounded by \eqref{eqn:Z_{t+1}_lower bound} and \eqref{eqn:Z_{t+1} upper bound}. It follows that $L(\Thetabf_{t})$ is bounded. As $T_o\rightarrow\infty$, we have $\frac{C'(\Gamma)(H_{0}-H_{T_o})}{T_o}\rightarrow0$ and $\frac{L(\Thetabf_{0})-L(\Thetabf_{T_o})}{VT_o}\rightarrow0$, and  \eqref{thm1:bd_longterm} follows.
\endIEEEproof

\section{Proof of Proposition~\ref{prop3}}\label{appE}
\IEEEproof
For $t=T_o$, from the dynamic shifting in \eqref{eqn:dynamic shift Z}, we have $Z_{T_o}=B_{T_0}-A_o-\Delta_a$. For $t=0$, we have $Z_{0}=B_{0}-A_o$. Thus, we have the following relation: $\frac{1}{T_o}(Z_{T_o}-Z_{0})=\frac{1}{T_o}(B_{T_o}-B_{0})-\frac{\Delta_a}{T_o}$. Substituting the first equation in \eqref{eqn:delta_a} into the above, we have
\begin{align*}
\frac{Z_{T_o}-Z_{0}}{T_o}=\frac{\sum_{\tau=0}^{T_o-1}[\eta_c(Q_\tau+S_{r,\tau})-\frac{F_{d,\tau}+F_{s,\tau}}{\eta_d}]-\Delta_a}{T_o}.
\end{align*}
Note that  $Z_t$ in \eqref{eqn:queue Z} is derived from  the above.
Since this finite time horizon algorithm, \eqref{eqn:delta_a} is satisfied with error $\epsilon=Z_{T_o}-Z_{0}$.
Because $Z_t$ is bounded by \eqref{eqn:Z_{t+1}_lower bound} and \eqref{eqn:Z_{t+1} upper bound} and $H_t$ is bounded by \eqref{eqn:app H bound}, the error $\epsilon$ has the following upper bound
\begin{align}
|\epsilon|
&\leq \max\{\eta_cH_{\max},|H_{\min}|/\eta_d-VP_{s}^{\min}\}+\eta_cR_{\max}\nn\\
&\quad +VP_{b}^{\max}-H_{\min}/\eta_d+D_{\max}/\eta_d\nn\\
&\leq \max\{H_{\max}/\eta_d+|H_{\min}|/\eta_d, 2|H_{\min}|/\eta_d-VP_{s}^{\min}\}\nn\\
&\hspace{2em}+VP_{b}^{\max}+\eta_cR_{\max}+D_{\max}/\eta_d. \nn \\
&=(2\Gamma+VC'(\Gamma))/\eta_d+\max\{0,VC'(\Gamma)/\eta_d-VP_{s}^{\min}\}\nn \\
&\quad+VP_{b}^{\max}+\eta_cR_{\max}+D_{\max}/\eta_d.
\end{align}
Thus, we complete the proof.
\endIEEEproof

\section{Proof of Proposition \ref{prop E times P zero}}\label{app E times P zero}
\IEEEproof
To ensure \eqref{equ:sell buy constraint} is satisfied, we must show the optimal control solution \eqref{case a}-\eqref{case f} in Proposition~\ref{prop0} can ensure \eqref{equ:sell buy constraint} being satisfied.
For Cases 1, 2 and 4, from their optimal control solutions \eqref{case a}, \eqref{case c} and \eqref{case f}, it is easy to see that \eqref{equ:sell buy constraint} is satisfied.
For Cases 3 and 5, from their optimal control solutions \eqref{case b 1} or \eqref{case b 2} and \eqref{case d 2} or \eqref{case d 3}, if $F^\w_{d,t}=W_t-S_{w,t}<D_{\max}$,  $E^\w_t=0$ and $F_{s,t}\geq0$; If $F^\w_{d,t}=D_{\max}$, we have $F^\w_{s,t}=0$ and $E^\w_t\geq0$.
If the battery is in the idle state, we always have $F_{s,t}^\textrm{id}=0$. Thus, \eqref{equ:sell buy constraint} is a sufficient condition for Algorithm~\ref{alg1}.
\endIEEEproof


\end{document}